\documentclass[preprint, superscriptaddress,notitlepage,
 amsmath,amssymb,
 aps, physrev,
]{revtex4-2}

\usepackage{hyperref}
\usepackage{graphicx}% Include figure files
\usepackage{dcolumn}% Align table columns on decimal point
\usepackage{bm}% bold math
\usepackage{color}
\usepackage{xcolor}
\usepackage{ulem}
\newcommand\tref[1]{~\ref{#1}}
\newcommand\tcite[1]{~\cite{#1}}

\begin{document}

\preprint{APS/123-QED}

\title{\textbf{Non-reciprocal coalescence-breakup dynamics in
    concentrated emulsions}}

\author{Ivan Girotto} \thanks{Equally contributed authors}
\affiliation{The Abdus Salam, International Centre for Theoretical
  Physics, Trieste, Italy} \author{Andrea Scagliarini} \thanks{Equally
  contributed authors} \affiliation{CNR-IAC, I-00185 Rome, Italy}
\affiliation{INFN, Sezione di Roma Tor Vergata, I-00133 Rome, Italy}
\author{Lei Yi} \thanks{Equally contributed authors}
\affiliation{Department of Physics, University of Massachusetts,
  Amherst, Massachusetts 01003, USA} \affiliation{New Cornerstone
  Science Laboratory, Center for Combustion Energy, Key Laboratory for
  Thermal Science and Power Engineering of MoE, Department of Energy
  and Power Engineering, Tsinghua University, 100084 Beijing, China}
\author{Roberto Benzi} \affiliation{Sino-Europe Complex Science
  Center, School of Mathematics, Northwestern University of China,
  Shanxi 030051, Taiyuan, China} \affiliation{Department of Physics
  and INFN, University of Rome Tor Vergata, Rome I-00133, Italy}
\author{Chao Sun} \email{Contact authors: FT: f.toschi@tue.nl; CS:
  chaosun@tsinghua.edu.cn; } \affiliation{New Cornerstone Science
  Laboratory, Center for Combustion Energy, Key Laboratory for Thermal
  Science and Power Engineering of MoE, Department of Energy and Power
  Engineering, Tsinghua University, 100084 Beijing, China}
\affiliation{Department of Engineering Mechanics, School of Aerospace
  Engineering, Tsinghua University, Beijing 100084, China}
\affiliation{Physics of Fluids Group, Max Planck-University of Twente
  Centre for Complex Fluid Dynamics, University of Twente, 7500 AE
  Enschede, The Netherlands} \author{Federico Toschi} \email{Contact
  authors: FT: f.toschi@tue.nl; CS: chaosun@tsinghua.edu.cn; }
\affiliation{Department of Applied Physics and Science Education,
  Eindhoven University of Technology, 5600 MB Eindhoven, The
  Netherlands} \affiliation{CNR-IAC, I-00185 Rome, Italy}

\date{\today}

\begin{abstract}
Dense stabilized emulsions are mixtures of immiscible fluids where the
high-volume fraction droplet dispersed phase is stabilized against
coalescence by steric interactions. The production of emulsions—a key
process in food, cosmetics and chemical industries—involves high-shear
flows, elastic and steric interactions, and proceeds thanks to
coalescence and breakup of droplets and interfaces. The complex
interplay between all these interactions is key in determining both
small-scale droplet morphology as well as large-scale emulsion
rheology. It is well known that at a critical volume fraction,
$\phi_c$, the emulsion loses stability, undergoing an extremely rapid
process where the fluid components in the emulsion exchange
roles. This process, called catastrophic phase inversion, which
resembles in several respects a dynamical phase transition, has
remained widely elusive from an experimental and theoretical point of
view. In this work, we present state-of-the-art experimental and
numerical data to support a dynamical-system framework capable of
precisely highlighting the dynamics occurring in the system as it
approaches the catastrophic phase inversion. Our study clearly
highlights that at high volume fractions, dynamical changes in the
emulsion morphology, due to coalescence and breakup of droplets, play
a critical role in determining emulsion’s rheology and
stability. Additionally, we show that at approaching the critical
volume fractions, the dynamics can be simplified as being controlled
by the dynamics of a correlation length represented, in our systems,
by the size of the largest droplet. This dynamics shares a close
connection with non-reciprocal phase transition where two different
physical mechanisms, coalescence and breakups, can get out of balance
leading to large non-symmetric periodic excursions in phase space. We
clarify the phenomenology observed and quantitatively explain the
essential aspect of the highly complex dynamics of stabilized
emulsions undergoing catastrophic phase inversion. More generally, our
approach sets the basis for the definition and modeling of a vast
number of dynamical phase transitions in hydrodynamic systems
out-of-equilibrium where the flow, or other advection mechanisms, can
enhance both aggregation and breakup of aggregates.
\end{abstract}

\maketitle

\noindent 
Many systems in nature are made of basic entities composing
dynamically evolving aggregates. The presence of external forces has a
dual role as these can considerably accelerate the aggregation process
and, at the same time, be responsible for the breakup of larger
aggregates.  This aggregation/breakup balance gives rise to a rich
dynamical equilibrium -typically characterized by a highly complex
statistics- that remains statistically stationary up to a critical
point. Beyond the critical point detailed balance is broken and the
system generically undergoes a phase transition with the formation of
a single integral-scale aggregate. While the dynamics in principle
depends on the multitude of physical properties characterizing the
system, the volume fraction of the aggregates and the intensity of the
external force play a key role in determining critical
points. Transitions of this kind include the colloidal sol-gel
transition in stirred vessels~\cite{Rouwhorst2020,Paradossi2002}, the
clusterization dynamics of floaters on turbulent
surfaces~\cite{Li2025}, including the formation of micro- and
macroplastics aggregates~\cite{Cozar2014,Wang2021}, planetesimals
formation where turbulence in the protoplanetary disk interferes with
accretion~\cite{Morbidelli2012}, collapse of
foams~\cite{yanagisawa2021dynamics,Wang201655} and the catastrophic
phase inversion in emulsions~\cite{perazzo2015phase,kumar2015recent}.

\begin{figure}[t!]
\centering
\includegraphics[width=0.95\textwidth]{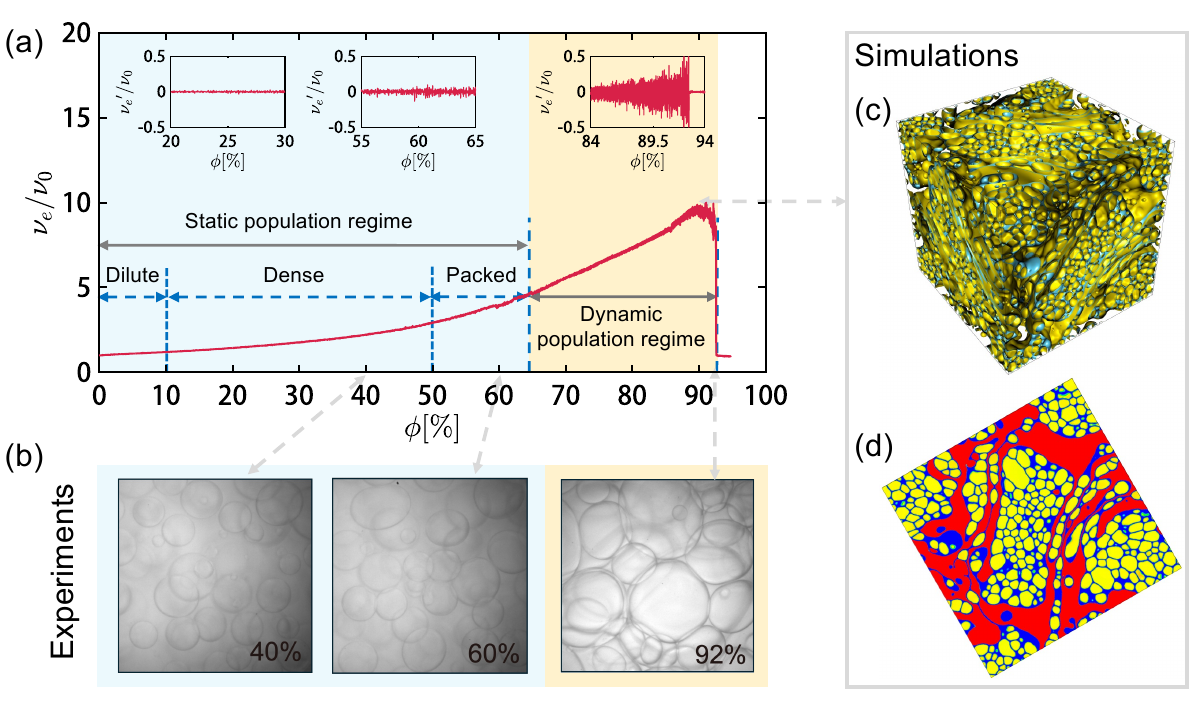}
\caption{Different regimes characterized by distinct rheology,
  dynamics, and morphology of the emulsion as the volume fraction of
  dispersed phase is slowly increased.  (a) For dilute volume
  fractions ($\phi \le 10\%$), the system effective viscosity
  $\nu_{e}$ increases almost linearly with $\phi$, with droplets
  remaining well-separated and showing minimal direct interactions. In
  the dense regime, the effective viscosity rises more steeply than
  linear as $\phi$ increases below $50\%$. When $\phi$ exceeds $50\%$,
  significant droplet deformation and enhanced effective viscosity
  $\nu_{e}$ are observed. The system remains in the static population
  regime for $\phi$ below $\sim 65\%$. For $\phi > 65\%$, the system
  transitions to a highly dynamic regime (dynamic population regime,
  DPR), marked by dramatically intensified fluctuations in the
  effective viscosity (${\nu_{e}}^{\prime}$ in the inset) and droplet
  size due to continuous coalescence and breakup events among
  densely-packed droplets.  Note that the effective viscosity is
  calculated based on the torque measured in a Taylor-Couette emulsion
  system, and $\nu_{0}$ is the kinematic viscosity of the water phase
  at $\phi=0\%$ (see \ref{sec:methods}).  (b) Snapshots from
  experiments at three volume fractions showing different
  morphological properties of the emulsion: $40\%$ (dense regime),
  $60\%$ (packed regime), and $92\%$ (DPR). (c,d) A 3D view and 2D
  cross-section of a snapshot in the dynamic population regime (DPR)
  from numerical simulations highlight the rich dynamics and droplet
  morphology ($\phi^{(\text{sim})}_c \approx 77\%$). The region
  colored in red indicates the volume occupied by the largest droplet.
}
\label{figure:firstpanel}
\end{figure}

\begin{figure}[t!]
\centering
\includegraphics[width=0.75\textwidth]{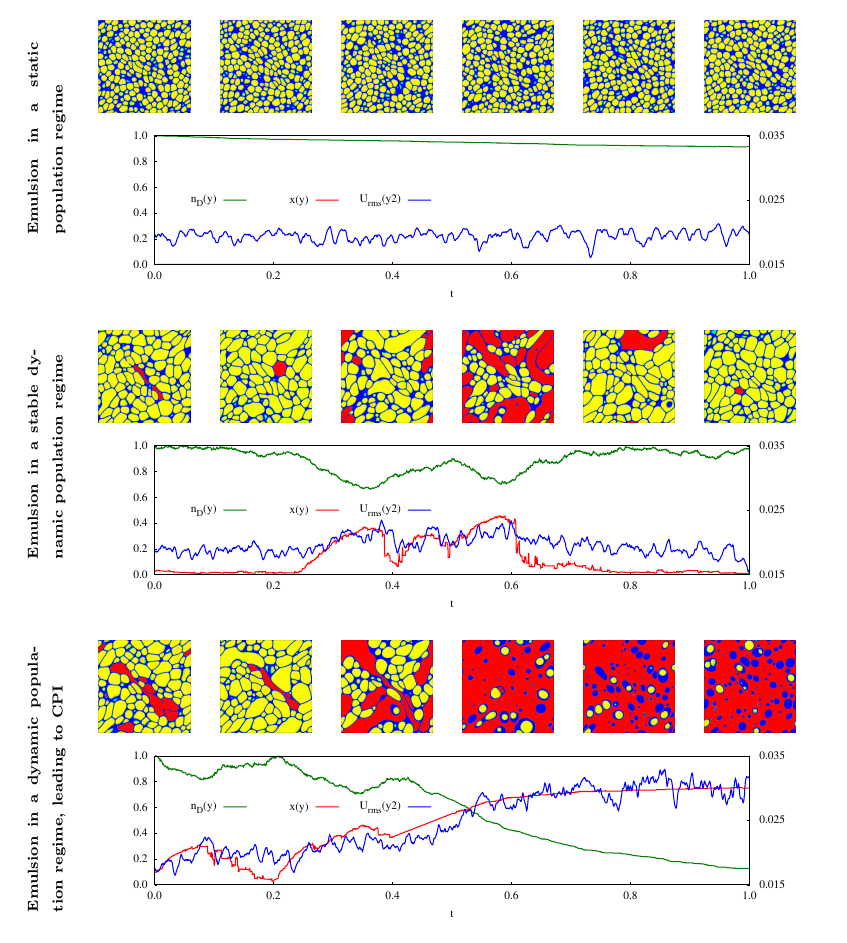}
\caption{Dynamics of emulsification processes via numerical
  simulations at constant volume fraction ($\phi^{(\text{sim})}$ =
  $38\%, 62\%$ and $77\%$) from packed (top) to highly packed
  (bottom). The number of droplets density, $n_{D}=N_{D}/N^{max}_{D}$,
  the volume of the largest droplet, $x$, and the mean square
  velocity, $U_{rms}$, are displayed as a function of
  time. Complemented by a series of 2-dimensional snapshots taken at
  the center of the 3-dimensional grid along the y-axes, from $t_{0}$
  and for five consecutive time intervals $t_{0}+\Delta(t)$, where
  $\Delta(t)=n\times0.2$ , ordered from left to right. The snapshots
  include a red region coloring the largest droplet in the emulsion
  (if present). The figure shows close to no dynamics for the packed
  emulsion $\phi^{(\text{sim})}=62\%$, while a significant activity is
  displayed when reaching the critical volume fraction of
  $\phi^{(\text{sim})}=77\%$.  In the last panel, the emulsification
  process shows the formation of a large droplet that suddenly and
  quickly leads to a CPI.}
\label{figure:pop_dynamics}
\end{figure}

\begin{figure}[t!]
\centering
\includegraphics[width=1.0\textwidth]{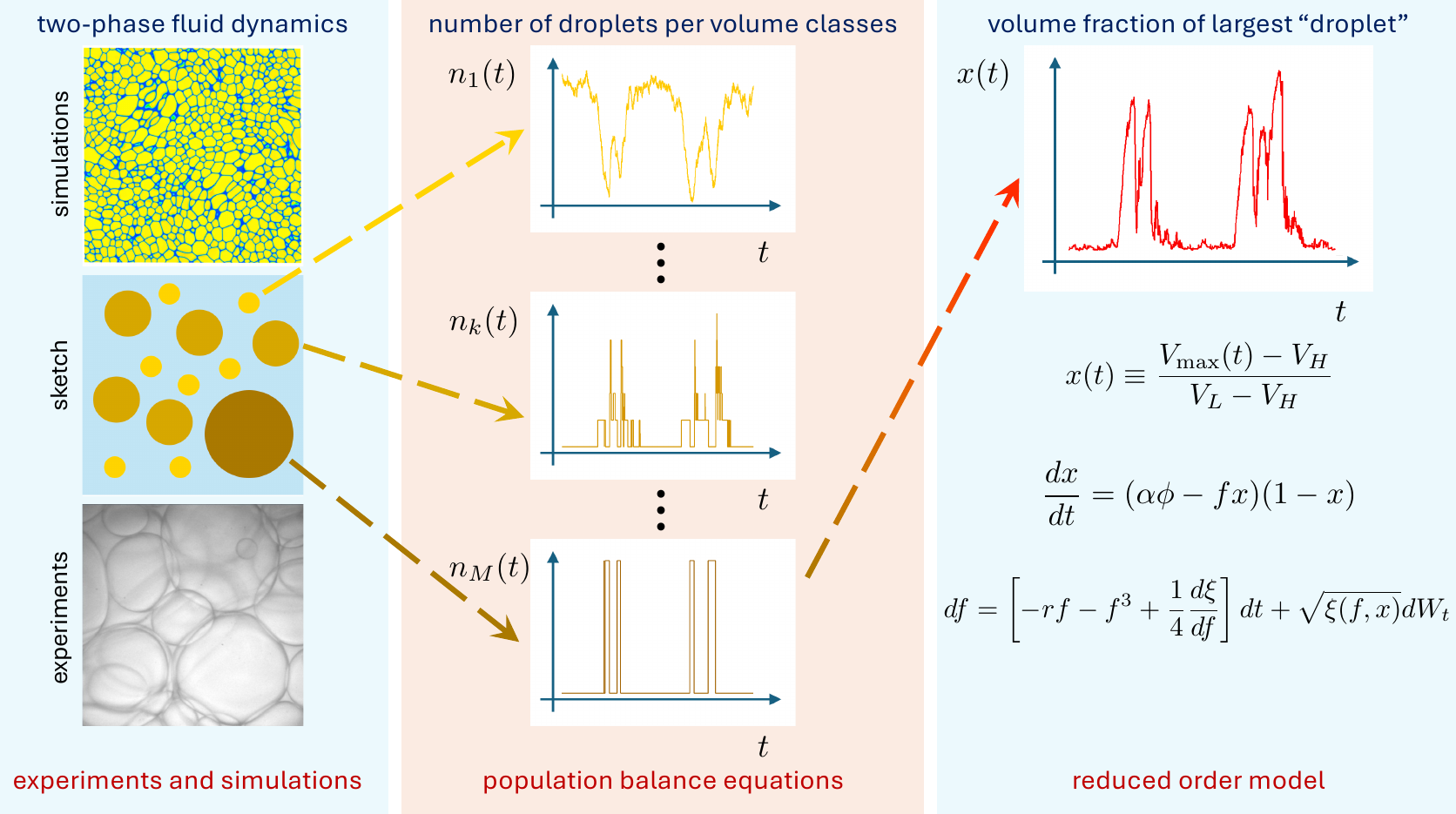}
\caption{Concept map of the model reduction steps leading to the
  identification of $x(t)$ (volume fraction occupied by the ``largest
  droplet'') as the relevant order parameter describing the dynamics
  in proximity of the catastrophic phase inversion and its evolution
  equation (see Eq. (\ref{eq:eq4x})). (Left panel) All the complexity
  of dynamical evolution of the full 3D hydrodynamical interaction
  between droplets, including surface tension and disjoining pressure,
  eventually including breakup and coalescence events is
  illustrated. (Central panel) In the dynamic population description
  the system is effectively considered as zero-dimensional and the
  dynamics variables representing the number of droplets of a given
  size display a non-trivial temporal dynamics only when breakup and
  coalescence events occur, all the complexity of the visco-elastic
  and hydrodynamic physics are neglected. (Right panel) At approaching
  the CPI only the dynamics of the largest droplet is relevant in
  order to describe the rheology. This dynamics can be captured by a
  simple set of two non-reciprocal ODEs, for $x(t)$ and for the
  breakup rate $f$, quantitatively modeling the statistics of the
  coalescence and breakup processes at varying the volume fraction
  $\phi$.}
\label{figure:sketch}
\end{figure}

\begin{figure}[t!]
\centering
\includegraphics[width=1.0\textwidth]{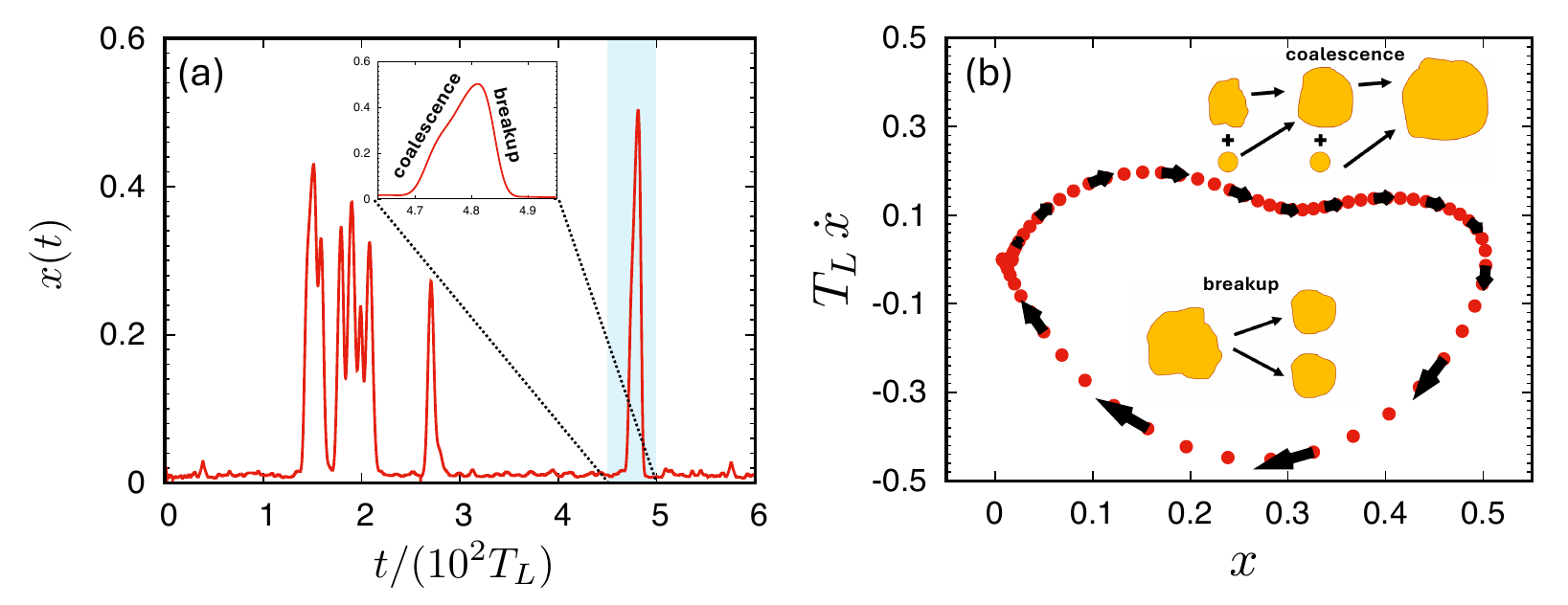}
\caption{(a) The signal of $x(t)$ as a function of time from a
  simulation at $\phi^{(\text{sim})} \approx 77\%$ ($T_L =
  L/U_{\text{rms}}$ is a characteristic large scale time); in the
  inset we show the same signal restricted to the time interval which
  we report in panel (b).  (b) Phase space portrait in the plane
  $(x,\dot{x})$ of the dynamics associated with a typical
  large-excursion, namely the one highlighted in panel (a) with a
  shaded area (and shown in the inset); the arrows indicate the
  forward time evolution.}
\label{figure:phase-portrait}
\end{figure}

Here, we focus on the latter example, as a prototype of the universal
phenomenon described above, that is amenable to controlled
experimental and numerical investigation.  Emulsions are soft
materials, widespread in biological and industrial
processes~\cite{leal2007emulsion}, consisting of a liquid-liquid
dispersion, stabilized by the presence of surfactants (see
Fig.\tref{figure:firstpanel}). At sufficiently high droplet volume
fraction, these systems are metastable and prone to phase inversion,
namely a microstructural, irreversible, change towards the
thermodynamically favoured configuration, whereby the continuous
minority phase becomes disperse (and vice versa droplets become the
continuous matrix)~\cite{perazzo2015phase,kumar2015recent}. The phase
inversion may occur in a smooth {\it transitional} way, typically when
it is triggered by changes in various parameters (such as temperature,
pH, chemical affinity), that affect the distribution of surfactant
over the phases\tcite{Vaessen1995,perazzo2015phase}. When induced by
increasing the concentration of disperse phase, the phase inversion is
sudden and abrupt, i.e. it is reminiscent of a {\it catastrophic}
process, in the sense of the theory of
catastrophes~\cite{Dickinson1981,Salager1988,Zeeman1976}.  Catastrophe
theory is essentially a bifurcation theory for the dynamics of a {\it
  behaviour} variable (related in some sense to the emulsion
morphology, in the case of phase inversion), occurring in the space of
control parameters which define the optimization function (a
generalized effective free energy)~\cite{Vaessen1995}. Despite some
success as a fitting method for experimental data, the catastrophe
theory approach to phase inversion suffers serious limitations,
specifically for what concerns its predictive capability and the
ambiguous physical interpretation of the morphology
parameters~\cite{Vaessen1996,Bouchama2003}.  Even more important, the
complex flow environment (turbulent or chaotic) under which
emulsification takes place does not enter in the description.  Yet,
wildly fluctuating hydrodynamic stresses in the stirring process lead
the emulsion out-of-equilibrium, tightly interwining mechanisms that
include turbulent dispersion, viscous stresses, elasticity,
hydrodynamics interactions with coalescence and breakup of droplets.
As a matter of fact, catastrophic phase inversion (CPI) in forced
systems still lacks a satisfactory theoretical interpretation.  Here,
leveraging state-of-the-art experiments and direct numerical
simulations (DNS), we probe the rheological and morphological
evolution of stirred emulsions.  Identifying the fraction of volume
occupied by the phase-inverted emulsion as the relevant {\it
  behaviour} (morphology) variable, we propose a stochastic dynamical
system, built on kinetic grounds and able to quantitatively capture
the statistical properties of the key empirical aspects.  Experiments,
simulations and the model provide agreeing evidence of clear
signatures of criticality, namely: i) the divergence of fluctuations
as a finite value of the volume fraction is approached and ii) the
emergence of non-Gaussian, bimodal, probability distribution
functions.  In particular, the model pinpoints the role played by the
intrinsic non-reciprocity of the breakup and coalescence processes,
and by the flow-induced effective mechanical noise in driving the
phase transition. \\
\noindent
{\bf Emulsification at high volume fraction} - Emulsification by
intense stirring of an immiscible mixture generates a turbulent
multiphase flow, whereby strong and fluctuating hydrodynamic stresses
break larger droplets into smaller droplets. In turn, when approaching
each other with sufficient kinetic energy to overcome disjoining
pressure barriers and lubrication forces, droplets coalesce minimizing
interfacial energy. This process leads to a dynamical equilibrium
characterized by a statistically stationary distribution of droplet
sizes (with a mean value typically compatible with the so-called
Kolmogorov-Hinze size
\cite{kolmogorov1949droplet,hinze1955fundamentals}).  To achieve a
concentrated (densely packed) emulsion of droplets
(e.g. ``oil-in-water'' or O/W emulsion), it is necessary to slowly add
the dispersed phase in order to maintain the system in an
intrinsically metastable state, i.e. energetically less favored with
respect to the inverted emulsion (e.g. ``water-in-oil'', W/O).
Further details on how this is realized in the experiments and in the
simulations are provided in the Materials and Methods section.\\ For
low-to-moderate volume fractions of the disperse phase, it is
customary to model the emulsification with population balance
equations for the droplet size distributions
\cite{Coulaloglou1977}. Under these not too high concentration
conditions, once the statistically stationary state is attained, the
emulsion morphology, or equivalently the droplet size distribution,
does not fluctuate much, i.e. the population balance regime is
relatively static; with reference to Fig.~\ref{figure:firstpanel}a,
this corresponds to the region $\phi \lesssim 65\%$ (hereafter $\phi$
will denote the volume fraction of oil).  In such static population
regime, the experimental results reported in
Fig.\tref{figure:firstpanel} show that, for dilute emulsions ($\phi
\le 10\%$), the system's effective viscosity increases nearly linearly
with $\phi$, with droplets remaining well-separated and showing
minimal direct interactions \cite{taylor1932viscosity}.  Note that the
system's effective viscosity $\nu_{e}$, obtained through time-resolved
global torque ($T$) measurements required to maintain a constant
angular velocity ($\omega_i$) of the inner cylinder in our TC system,
is given by $\nu_{e}/\nu_{0}=(T/T_{0})^{2.4}$. Here, $\nu_{0}$ and
$T_{0}$ are the viscosity of the continuous phase ($\phi=0\%$) and the
torque of the system with $\phi=0\%$ at the same rotational angular
velocity, respectively (see Section~\ref{subsec:exp} for
details).\\ In dense conditions ($10\% \lesssim \phi \leq 50\%$), the
effective viscosity rises more steeply than linear and above
$\phi=50\%$, one observes significant droplet deformation and enhanced
effective viscosity. When $\phi > 65\%$, the system transitions to a
highly dynamic population regime characterized by dramatically
intensified fluctuations in both effective viscosity (see top insets
of Fig.\tref{figure:firstpanel}) and droplet size, resulting from
continuous coalescence and breakup events among densely-packed
droplets.  The value of volume fraction ($\phi \approx 65\%$)
discriminating between static and dynamic population regimes
corresponds, approximately, to the random close packing of spheres in
three dimensions. Interestingly, as it was found in our numerical
simulations (see Fig. 3 and Table 1 in \cite{girotto2024lagrangian}),
it is also close to the value of $\phi$ at which the characteristic
mean `life-time' of droplets, defined as $t_D = \langle N \rangle
/\langle \beta \rangle$ (where $\langle N \rangle$ is the mean number
of droplet and $\langle \beta \rangle$ the mean breakup rate), becomes
of the order of the large eddy turnover time, suggesting that the
hydrodynamic flows and the interfacial dynamics start to get tightly
coupled.  While the dynamic population regime should theoretically
exhibit higher effective viscosity (and consequently lower effective
Reynolds number and thus lower turbulent fluctuations), we
surprisingly find extreme fluctuations in both the effective viscosity
(insets in Fig.\ref{figure:firstpanel}a) and droplet size variations
from experiments (Fig.\ref{figure:firstpanel}b) and numerical
simulations (Fig.\ref{figure:firstpanel}d).  At a critical volume
fraction $\phi_c$, in the dynamic population regime, the metastable
emulsion undergoes a catastrophic phase inversion, suddenly passing
from a concentrated oil-in-water dispersion to a dilute water-in-oil
dispersion (O/W $\rightarrow$ W/O). Correspondingly, the effective
viscosity drops abruptly to a low value, see
Fig.~\ref{figure:firstpanel}a, compatible with that of a dilute
emulsion at $\phi_{W/O} = 1-\phi_c$.  It must be remarked that the
actual value of $\phi_c$ depends on the flow conditions and on the
physico-chemical properties of the interface (encoded in the
dimensionless Reynolds, $Re$, and Weber, $We$,
numbers)\tcite{Vaessen1995,perazzo2015phase}; in our experiments
$\phi^{(\text{exp})}_c \approx 92\%$, while in the simulations
$\phi^{(\text{sim})}_c \approx 77\%$. The value of $\phi_c$ may also
depend on the system properties and turbulence level
\cite{bakhuis2021catastrophic,piela2009phenomenological}, moreover it
is worth stressing that, due to the diffuse interface nature of the
numerical model, if a quantitative matching between experiments and
simulations is sought for, a corrected effective volume fraction,
taking into account the finite interface width, should be considered,
as discussed in the Materials and Methods sections. Therefore, in what
follows we will always refer to $\phi^{(\text{exp})}$ and
$\phi^{(\text{sim})}$ for the volume fractions in experiments and DNS,
respectively, whenever actual numerical values are
involved.\\ Fig.~\ref{figure:pop_dynamics} helps us in getting
insights on the emulsion morphology, whose evolution supports the
distinction between the two population regimes below and above
$\phi^{(\text{exp})} \approx 65\%$ (arguably also this value, likewise
$\phi_c$, is expected to be $Re$ and $We$ dependent).  Numerical
snapshots of the oil density field at various instants of times, while
the stirring force is active and the volume fraction is kept constant,
are reported in Fig.~\ref{figure:pop_dynamics} for, from top to
bottom, $\phi^{(\text{sim})} \approx 38\%, 62\%$ and $77\%$. Oil
regions are coloured in yellow, water regions in blue, and the largest
oil-connected region (the ``largest droplet'') is coloured in red.
Each set of snapshots is shown together with the time evolution of the
corresponding number of oil droplets ($N_D(t)$, divided by its initial
value, $N_D(t_0)$), the volume of the largest oil droplet (normalized
by the total volume), $x(t)$, and the root mean square velocity
($U_{\text{rms}}$).  We observe that, for $\phi^{(\text{sim})} = 38\%,
62\%$ (i.e. in the static population regime), both $N_D(t)$ and $x(t)$
are almost constant in time, suggesting that the interfacial dynamics
(droplet breakups/coalescences) is unimportant. Accordingly, in this
regime, also $U_{\text{rms}}$ does not fluctuate significantly. At
$\phi^{(\text{sim})} = 77\%$ (i.e. in the so-called dynamic population
regime), instead, breakups and coalescences occur continuously in
time, on average balancing each other and establishing a dynamical
equilibrium. Such detailed balance can be occasionally and transiently
broken giving rise to very large fluctuations where the largest
droplet can occupy almost half of the total volume ($x(t) \sim
0.4$). These events of extreme transient growth and shrinkage of
$x(t)$ ({\it excursions}) come along with an important variation of
the $U_{\text{rms}}$, reflecting the fluidization of the part of the
system that is locally phase-inverted.  Catastrophic phase inversion
occurs when the largest oil droplet invades the whole volume,
eventually leading to the dispersion of water droplets in oil (last
two rows of Fig.~\ref{figure:pop_dynamics}) and, correspondingly,
$x(t) \sim 1 - \phi$.  Fig.~\ref{figure:firstpanel}(d) and
Fig.~\ref{figure:pop_dynamics} also highlight that the inversion goes
through states whereby regions of concentrated and phase-inverted
emulsions coexist (double emulsions). We recall that phase coexistence
in equilibrium systems is a trademark of discontinuous phase
transitions \cite{Binder1987}.
The physics of the dynamic population regime-characterized by emulsion
morphology and effective-viscosity fluctuations-is the central focus
of this work and is discussed in detail in the following sections.\\
\noindent 
{\bf Droplet population dynamics at approaching CPI} - The fraction of
volume occupied by the largest oil droplet, $x(t)$, features certain
properties that make it a suitable candidate as a dynamical order
parameter to probe the CPI transition. In the limit of large system
volumes, $x(t)$ goes from $x\sim 0$ (concentrated emulsion) to $x\sim
1-\phi$ (phase-inverted emulsion) across the transition. At volume
fractions close to CPI, $\phi \lesssim \phi_c$, it develops large
fluctuations (see Fig.~\ref{figure:pop_dynamics} and
Fig.~\ref{figure:phase-portrait}(a)), reminiscent of what was observed
experimentally for the effective viscosity
(Fig.~\ref{figure:firstpanel}).  This suggests that $x(t)$ may also
work as a proxy for the rheological response of the material.  In
fact, in the proximity of the phase inversion and from a rheological
point of view, the system can be seen as constituted of regions
occupied by the concentrated emulsion (a yield-stress, non-Newtonian
material with a relatively higher viscosity) and regions occupied by
the phase-inverted emulsion (a Newtonian fluid with a relatively lower
viscosity).  The idea of identifying $x$ as the relevant order
parameter for the CPI transition is appealing, but deserves a more
constructive justification. To this aim one may follow a conceptual
path that echoes the usual model reduction approach leading from the
atomistic view to the macroscopic continuum dynamics, which is
pictorially represented in Fig.~\ref{figure:sketch}.
The steady state of stirred non-dilute emulsions is a complex tangle
of droplet deformation, breakup and coalescence, driven by
hydrodynamic (turbulent) stresses, fluctuating in space and time; in a
stable emulsion, on average breakup and coalescence balance,
eventually determining a statistical distribution of droplet sizes
peaked around a mean value (in general much smaller than the system
size). From the perspective of emulsion morphology, this extremely
complex spatiotemporal dynamics can be reduced to a set of balance
equations for droplet populations (probability distributions),
$n_i(t)$ on the discrete space of size
classes~\cite{Coulaloglou1977,Krapivsky2010,Maass2012,Brilliantov2015}
(see also \ref{subsec:pbm} for more details). We posit that, at
approaching CPI, this ``kinetic''-level description might be further
simplified, looking just at the dynamics of the volume $M(t)$ of the
largest droplet. It is the latter, in fact, that discriminates, by
definition (it corresponds to the phase-inverted volume), when the
phase-transition takes place.  The whole hierarchy of equations for
the various $n_i$, with $i<M$, determines the effective in- and
out-fluxes for the largest droplet, i.e., the growth or decrease rates
for the dynamics of $x$.\\
\noindent {\bf Non reciprocal phase transition} - The dynamics of
$x(t)$ should be described by an ordinary differential equation
featuring source and sink terms. The source term stems from the
coalescence of smaller droplets (i.e. the input from the full
population balance hierarchy), whereas the sink is due to the largest
droplet breakup (into smaller droplets).  A key empirical evidence
regarding the largest droplet dynamics is that there is a fundamental
asymmetry between coalescence and breakup: while coalescence-driven
growth of $x$ occurs by progressive absorption of small drops of
normalized volume $\delta$ ($x\rightarrow x+\delta$), breakup is
characterized by abrupt, mostly binary, events ($x \rightarrow x/2$).
This breakdown of detailed balance in the transitions between
microscopic configurations causes the reported transient excursions in
the signal of $x(t)$ (Fig.~\ref{figure:phase-portrait}(a)). The
associated phase portrait, plotted in
Fig.~\ref{figure:phase-portrait}(b), is represented by a closed curve
in the $(x,\dot{x})$ plane, whose asymmetry with respect to the
$\dot{x}=0$ axis confirms the microscopic detail balance breakdown
and, consequently, suggests that the two processes, coalescence-driven
growth and breakup-driven shrinkage of the largest droplet, are
characterized by distinct timescales.\\ Since the coalescence rate is
expected to be an increasing function of the volume fraction $\phi$, a
minimal dynamical model for $x$ can be written down as:
\begin{equation}\label{eq:eq4x}
    \frac{dx(t)}{dt} = [\alpha \phi -f(t)x(t)]  [1-x(t)],
\end{equation}
where $\alpha$ is a constant (a characteristic rate of coalescence),
$f$ can be interpreted as the breakup rate and the overall factor
$(1-x)$ has been introduced since for $x=1$ (i.e. the whole volume
undergoes phase inversion) the process must stop.  Figures
\ref{figure:pop_dynamics} and \ref{figure:phase-portrait}(a) tell that
the dynamics of $x(t)$ displays bursting events where $x$ reaches
large values ($x \sim 0.5$), spaced out by relatively long periods
where it remains small ($x \sim 0$).  Modelling the ratio $\alpha/f$,
then, needs to take into account such observation.  To this aim, we
assume $\alpha$ to be constant and $f$ to satisfy the stochastic
differential equation:
\begin{equation}\label{eq:eq4f}
df = \left[-r f - \beta f^3 + \frac{1}{4} \frac{d\xi(f,x)}{df}\right]dt + \sqrt{\xi(f,x)} dW_t,
\end{equation}
where $W_t$ is the standard Wiener's process and $r^{-1}$ is a
characteristic time for the fragmentation rate to vanish. The
stochastic dynamics of $f$, as determined by the above equation,
accounts for the effect of the fluid fluctuating turbulent stresses
acting on the droplet interfaces and inducing them to break up. This
effect is embedded in the variance of the multiplicative noise,
$\xi(f,x)=\sqrt{\varepsilon_0 + \varepsilon_1 f^2(1-x)}$. Further
details and theoretical justification of
Eqs.~(\ref{eq:eq4x})-(\ref{eq:eq4f}) can be found in
\ref{subsec:model}.\\
%%%%%%%%%%%%%%%%%%%%%%%%%%%%%%%%%%%%%%%%%%%%%%%%%%%%%%%%%%%%%%%%%%%%%%%%
\begin{figure}[t!]
\centering
\includegraphics[width=0.8\textwidth]{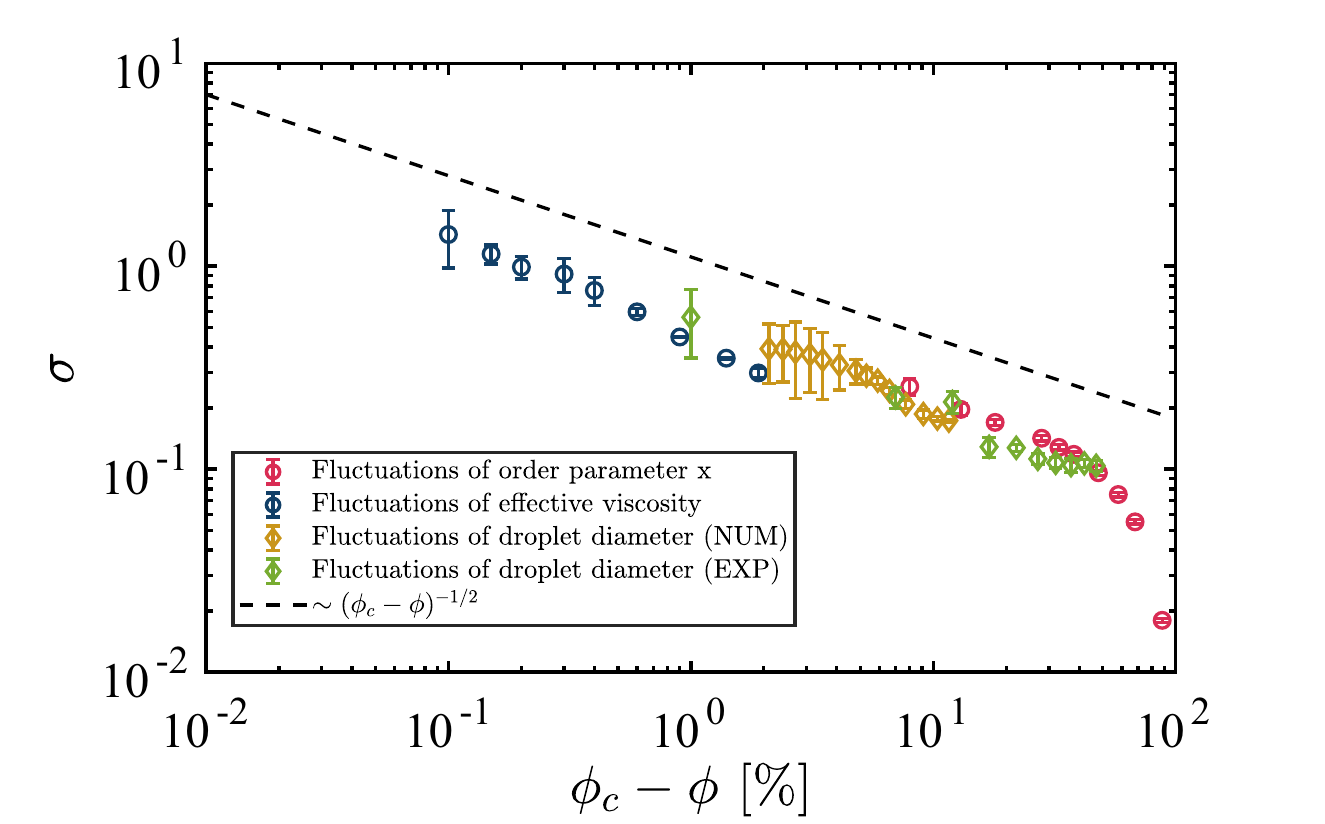}
\caption{Fluctuations (standard deviation) of the order parameter, the
  effective viscosity, droplet diameter from experiments (EXP), and
  droplet diameter from numerics (NUM) as functions of $(\phi_c -
  \phi)$, with $\phi_c = 92\%$ (EXP) and $\phi_c = 77\%$ (NUM).
  Fluctuation data of the effective viscosity and droplet diameters
  (both NUM and EXP) are normalized with given constant values for
  comparison among different quantities here.  The dashed line
  highlights the power law divergence $\sigma \sim (\phi_c -
  \phi)^{-1/2}$.  }
\label{figure:fluct_x}
\end{figure}
%%%%%%%%%%%%%%%%%%%%%%%%%%%%%%%%%%%%%%%%%%%%%%%%%%%%%%%%%%%%%%%%%%%%%%%
In Fig.~\ref{figure:fluct_x} we plot the fluctuations of $x$
(expressed by the standard deviation $\sigma = \sqrt{\langle x^2
  \rangle - \langle x \rangle^2}$, where the average is meant over
time and over $O(10^2)$ realizations of the noise) obtained from the
numerical integration of the model
Eqs.(\ref{eq:eq4x})-(\ref{eq:eq4f}), together with the effective
viscosity fluctuations measured in the experiments.  In all cases, we
observe a divergent behaviour of the $\sigma$ as a function of the
distance from the CPI transition point $\phi_c$ as $\sigma \sim
(\phi_c - \phi)^{-1/2}$.  It must be noted, as for any phase
transition in a finite system, divergences will have a cutoff,
specifically the variance of $x$ cannot exceed unity and a real
divergence would be expected only in the thermodynamic (infinite
volume) limit.\\ Our study highlights a certain analogy between CPI
and non-reciprocal phase transitions \cite{Fruchart2021}, namely the
asymmetric role of breakup and coalescence in the droplet population
dynamics (causing the breakdown of microscopic detailed balance) and,
at a more mathematical level, the non-hermitianity of the Jacobian of
the deterministic version ($W_t=0$ identically) of the system
(\ref{eq:eq4x})-(\ref{eq:eq4f}) \cite{Fruchart2021}.

\begin{figure}[t!]
\centering
\includegraphics[width=1.0\textwidth]{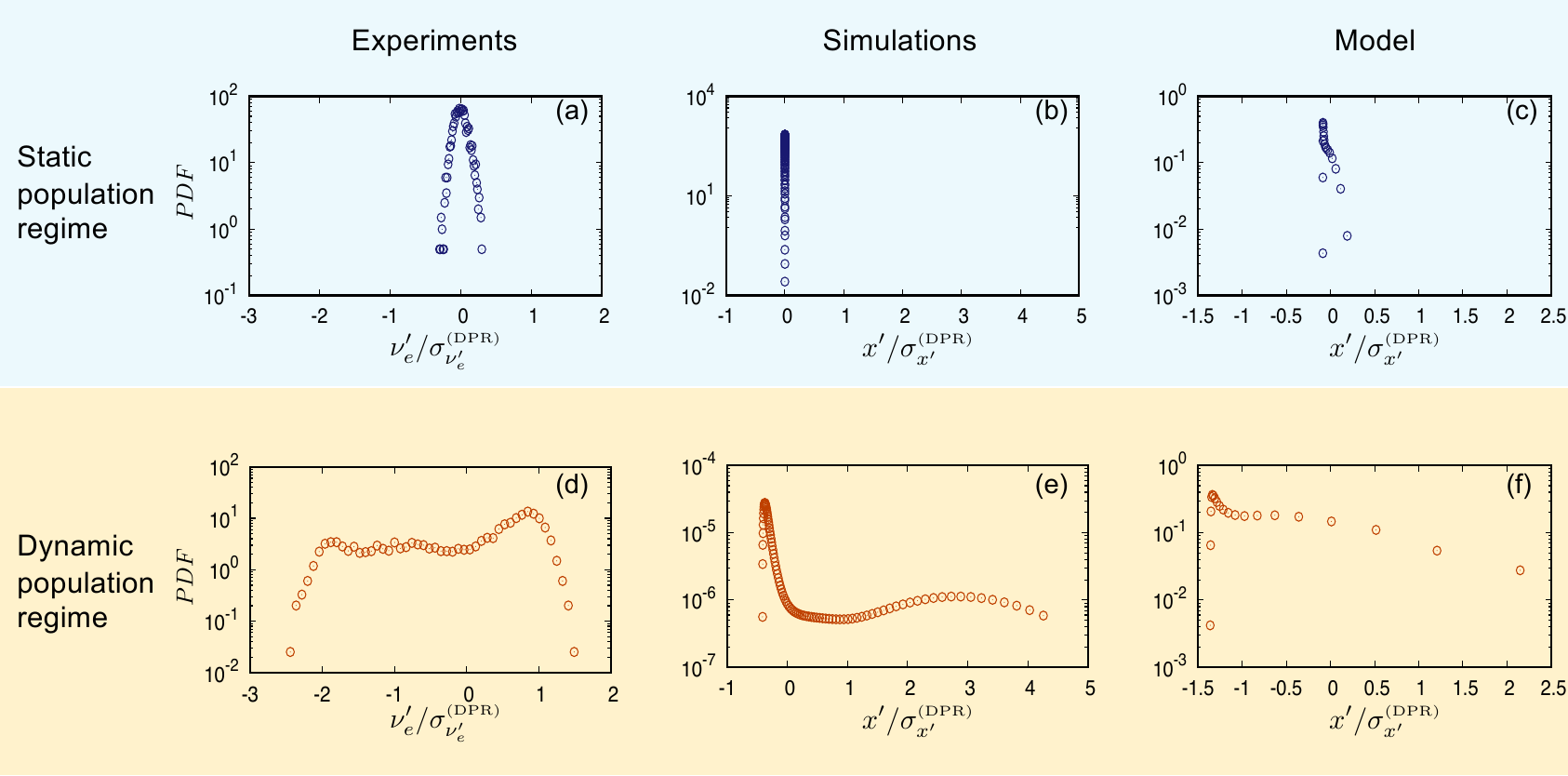}
\caption{PDFs of effective viscosity fluctuations ${\nu_{e}}^{\prime}$
  from experiments (a,d), and PDFs of $x^{\prime}=x-\langle x
  \rangle$, fluctuations of the fraction of total volume occupied by
  the phase-inverted emulsion, from the direct numerical simulations
  (b,e) and from the model (c,f), at moderate volume fraction (in the
  'static population regime', (a,b,c)) and high volume fraction (in
  the 'dynamic population regime', (d,e,f)). All quantities are given
  in units of their standard deviations in the dynamic population
  regime ($\sigma_{\nu^{\prime}}^{\mbox{\tiny{(DPR)}}}$ and
  $\sigma_{x^{\prime}}^{\mbox{\tiny{(DPR)}}}$, respectively).  Notice
  the bimodality in the dynamic population regime, with the secondary
  peak (at lower ${\nu_{e}}^{\prime}/\nu_{0}$, high $x$) originating
  from transient states where concentrated and phase-inverted emulsion
  coexist.}
\label{figure:PDFS}
\end{figure}

%%%%%%%%%%%%%%%%%%%%%%%%%%%%%%%%%%%%%%%%%%%%%%%%%%%%%%%%%%%%%%%
\noindent {\bf Bridging experiments, simulations and theory together:
  statistics of critical fluctuations} - We measured the temporal
fluctuations of the effective viscosity (applied torque) in a
Taylor-Couette experiment, as the volume fraction of initially
dispersed phase is increased while keeping the angular velocity of the
inner cylinder constant (further details can be found in
\ref{sec:methods}). Close to the phase-inversion point, the average
effective viscosity slightly decreases while the fluctuation is
progressively growing.  At some volume fractions, we fix $\phi$ and
perform long-time measurements (step-by-step) to investigate the
critical dynamics.  Note that the step duration $\tau_{\text{exp}} =
60$ min is $\mathcal{O}(10^{6})$ times the turnover time scale of the
flow.  The most relevant aspect to be highlighted is that the
probability density function (PDF) of the fluctuations of effective
visocisty, ${\nu_{e}}^{\prime}$, which is Gaussian and narrow in the
static population regime (dilute emulsion, Fig.~\ref{figure:PDFS}(a)),
tends to develop a non-Gaussian bimodal shape as the phase inversion
is approached in the dynamic population regime
(Fig.~\ref{figure:PDFS}(d)).  This suggests that the system spends
most of the time in the concentrated emulsion state (main peak), but
visits, with non-negligible frequency, the phase coexistence state
discussed above (secondary peak at lower effective viscosity
fluctuations $\nu_e'$).\\ As previously discussed, we do expect the
statistics of the effective visocity, $\nu(t)$, to be closely related
to that of the largest droplet volume, $x(t)$.  This is confirmed,
indeed, in Fig.~\ref{figure:PDFS}(b,e), where we plot the PDFs of the
fluctuations $x^{\prime} = x - \langle x \rangle$ obtained from the
numerical simulations, in the two regimes.  The PDFs share strong
similarities with the experimental PDFs of ${\nu_{e}}^{\prime}$:
notice, in particular, the presence, in the dynamic population regime,
of the two peaks, with the secondary one associated to the occurrence
of a partial phase inversion.  Remarkably, the emergent bimodality as
$\phi$ increases can be detected also in the PDFs of $x^{\prime}$
computed from the numerical integration of the model
Eqs.~(\ref{eq:eq4x})-(\ref{eq:eq4f}) and reported in
Fig.~\ref{figure:PDFS}(c,f).\\
%%%%%%%%%%%%%%%%%%%%%%%%%%%%%%%%%%%%%%%%%%%%%%%%%%%%%%%%%%%%%%%%%%%
%%%%%%%%%%%%%%%%%%%%%%%%%%%%%%%%%%%%%%%%%%%%%%%%%%%%%%%%%%%%%%%%%%%%%%

{\bf Conclusions and perspectives}\\ A large number of systems in
nature are made of elementary entities which aggregate and break up
under the influence of external forces such as, for example,
hydrodynamic drag. The aggregates can, to some degree, oppose these
external forces depending on their size and physical structure, giving
rise to a complex dynamics of aggregation / breakup which becomes more
and more intermittent at increasing the volume fraction. As a specific
case, we focus on stabilized emulsions of immiscible fluids where the
aggregation process consists in droplet coalescence and the breakup of
droplets into smaller droplets. While the flow determines the
coalescence and breakup rates, the viscoelastic interaction between
droplets influences the flow itself by determining its rheology. In
this work we introduced a theoretical framework that allows capturing
the key phenomenology for very large concentrations, where the
dynamics of coalescence and breakup is particularly active, and we
show that the macroscopic rheological properties of the system are
dynamically and precisely determined by the population dynamics
between droplets of different sizes. Additionally, the theoretical
framework introduced allows us to quantitatively define the concept of
a dynamical order parameter and dynamics phase transition for this
type of systems.

\begin{figure}[t!]
\centering \includegraphics[width=1\textwidth]{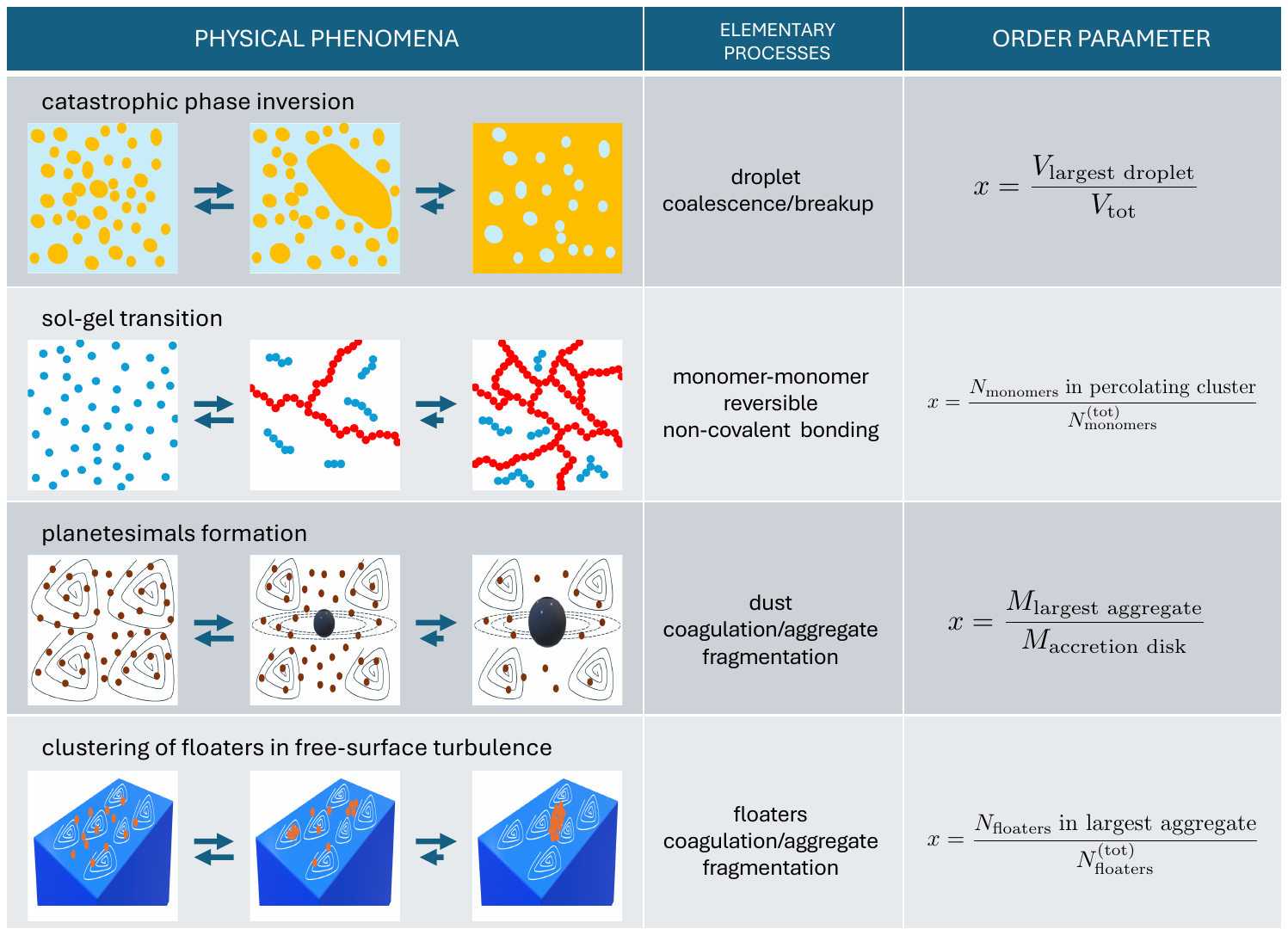}
\caption{Some instances of physical systems whose dynamics is
  characterized by aggregation-fragmentation of elementary units in
  turbulent flow environments and such that a non-equilibrium phase
  transition with the formation of a single large scale aggregate can
  take place. From top to bottom: catastrophic phase inversion in
  emulsions; sol-gel transition in stirred vessels; planetesimal
  formation in a turbulent proto-planetary disk; clustering of
  floaters in free-surface turbulent flows (e.g. micro-/macro-plastics
  on the ocean surface).}
\label{figure:applications}
\end{figure}

Owing to the generality of the theoretical approach, our study offers
a solid statistical physics framework to explain the basic physics
beyond CPI in emulsions, that -more generally- embraces all those
systems undergoing a phase transitions characterised by broken
detailed balance of small-scale aggregation-fragmentation processes (a
few instances are reported and set in our theoretical framework in
Fig.~\ref{figure:applications}).
%%%%%%%%%%%%%%%%%%%%%%%%
%%%%%%%%%%%%%%%%%%%%%%%%%%

\section{Materials and Methods}\label{sec:methods}

\subsection{Experiments}\label{subsec:exp}
\begin{figure}[htbp]
\centering \includegraphics[width=0.9\textwidth]{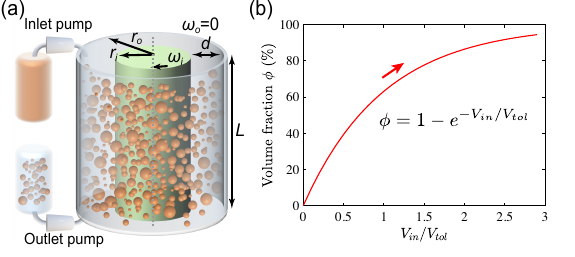}
\caption{ (a) A sketch of the experimental setup. The emulsion was
  maintained in the gap by rotating the inner cylinder at a constant
  angular velocity $\omega_{i}$. The torque sensor was used to measure
  the torque exerted on the inner cylinder with high accuracy, which
  is used to calculate the effective viscosity of the emulsion.  Two
  micropumps were used to gradually change the volume fraction $\phi$
  of the dispersed phase.  (b) The volume fraction of the dispersed
  phase in the emulsion versus $V_{in}/V_{tol}$.  }
\label{fig_SM_EXP_setup}
\end{figure}

%%%%%%%%%%%%%%%%%%%%%%%%%%%%%
The experiments are performed in a Taylor-Couette (TC) turbulence
system (Fig.~\ref{fig_SM_EXP_setup}(a)), which is the flow confined
between two coaxial cylinders.  The TC system is characterized by an
outer cylinder of radius $r_{o} = 35\rm~mm$, an inner cylinder of
radius $r_{i} = 25\rm~mm$, a gap $d = r_{o} - r_{i} = 10\rm~mm$, and a
height $L = 75\rm~mm$.  The emulsion confined between the two
cylinders contains two immiscible liquids: silicone oil (density
$\rho_{o}=866\rm~kg/m^{3}$, viscosity $\nu_{o}=2.1 \times
10^{-6}\rm~m^{2}/s$) from Shin-Etsu and an aqueous ethanol-water
mixture ($75\%$ volume fraction of water phase,
$\rho_{w}=860\rm~kg/m^{3}$, $\nu_{w}=2.4\times10^{-6}\rm~m^{2}/s$).
Because the oil and ethanol-water mixture is inherently unstable, we
maintain it in dynamic equilibrium using Taylor-Couette turbulent
flow~\cite{Grossmann2016}, which continuously mixes the emulsion.
This requires constant energy input through the rotation of the inner
cylinder to sustain the turbulent state (the outer cylinder is
stationary).  Note that the densities of the two liquids are almost
matched, eliminating the effect of the centrifugal force.

In the experimental procedure, we gradually injected the dispersed
phase (either oil or water) into the system at a constant volume flow
rate $Q$ to slowly increase its volume fraction $\phi$ (a quasi-static
process, see Fig.~\ref{fig_SM_EXP_setup}(b)).  During injection, we
determined the instantaneous oil volume fraction by assuming a uniform
distribution of the injected liquid throughout the emulsion due to
turbulent mixing - an assumption that was experimentally verified
(Fig.~\ref{fig_SM_EXP_setup}(b)).  As the total volume of the liquid
$V_\text{tol}$ and the injection flow rate $Q$ are known, the
dispersed-phase volume fraction as a function of time (or injected
liquid volume $V_\text{in}$) can be analytically derived as:
\begin{equation}
\phi = 1 - e^{-\frac{V_\text{in}}{V_\text{tol}}} =1 - e^{-\frac{Q}{V_\text{tol}} t},
\label{equ1}
\end{equation}
where $t$ is the time from the start of injection.  We perform two
types of measurements on the flowing emulsion.  The first one is the
system's effective viscosity $\nu_{e}$ obtained through time-resolved
global torque ($T$) measurements required to maintain a constant
angular velocity ($\omega_i$) of the inner cylinder in our TC system
(Fig.~\ref{figure:firstpanel}a).  Note that the effective viscosity is
calculated by $\nu_{e}/\nu_{0}=(T/T_{0})^{2.4}$, where $\nu_{0}$ and
$T_{0}$ are the viscosity of the pure water phase ($\phi=0\%$) and the
torque of the system with $\phi=0\%$ at the same rotation angular
velocity.  This calculation is based on the assumption that the
momentum transport in the TC system follows the same rule for
different volume fractions and Reynolds numbers (more details can be
found in ~\citep{yi2021global}).  The second measurement is the local
evolution of droplet structures, including their size and morphology
(Fig.~\ref{figure:firstpanel}b).  Additionally, at some volume
fractions near the phase-inversion point, we fix the $\phi$ and
perform long-time measurements (step-by-step) to investigate the
critical dynamics.  Note that the step duration $\Delta t = 60$ mins
is $\mathcal{O}(10^{6})$ times the turnover time scale of the flow.
%%%%%%%%%%%%%%%%%%%%%%%%%%%%%%

\subsection{Numerical simulations}\label{subsec:sim}
The simulations are performed integrating numerically, by means of a
two-component lattice Boltzmann method
(LBM)\cite{BENZI1992145,succi2018lattice}, the equations of motion for
the two fluid density fields, indicated as $O$ and $W$ (for, e.g.,
`oil' and `water'), $\rho_{O,W}$, and the incompressible barycentric
fluid velocity field, $\mathbf{u}$:
\begin{align}\label{eq:eom}
    \partial_t \rho_{\sigma} + \nabla \cdot (\rho_{\sigma}\mathbf{u})
    & = 0 \quad \sigma = O,W \nonumber \\ \rho_f (\partial_t
    \mathbf{u} + \mathbf{u} \cdot \nabla \mathbf{u}) & = - \nabla p +
    \nabla \cdot \mathbf{P}^{\text{(mix)}} + \eta \nabla^2 \mathbf{u}
    + \mathbf{F}^{(\text{ext})},
\end{align}
where $\rho_f = \rho_O + \rho_W$ is the total density, $p$ is the
pressure, $\eta$ is the dynamic viscosity and
$\mathbf{F}^{(\text{ext})}$ a forcing
term. $P^{(\text{mix})}_{ij}[\rho_O, \rho_W, \partial_i \rho_O,
  \partial_i \rho_W,\dots]$, a function of the two density fields and
of their derivatives, is the non-ideal contribution to the pressure
tensor.  In this hydrodynamic framework, the surface tension,
$\gamma$, is computed from its mechanical definition as the integral
of the mismatch between the normal and tangential components of
$\mathbf{P}^{(\text{mix})}$ across a flat interface at equilibrium,
i.e., considering without loss of generality a $2d$ problem, $\gamma =
\int_{-\infty}^{+\infty} \left(P_{xx}^{(\text{mix})}
-P_{yy}^{(\text{mix})}\right)(x) dx$~\cite{rowlinson2002molecular}.
Analogously, the disjoining pressure, $\Pi$, can be determined by
measuring the overall film tension, $\Gamma_f$, as the integral of the
normal-transversal stress tensor mismatch across two flat interfaces
separated by a liquid layer of width $h$. The disjoining pressure is
related to $\Gamma_f$ as $h\frac{d \Pi}{d h} = \frac{d
  \Gamma_f(h)}{dh}$
\citep{bergeron1999forces,sbragaglia2012supramolecular}.\\ The
large-scale forcing needed to generate the chaotic stirring that mixes
the two fluids, $\mathbf{F}^{\text{(ext)}}$, takes the following form:
\begin{equation}
  F^{\text{(ext)}}_i(\mathbf{x}, t) = \sum_{\sigma}
  F^{\text{(ext)}}_{\sigma i}(\mathbf{x}, t) = A \sum_{\sigma}
  \rho_{\sigma} \sum_{k \leq 2\pi \sqrt{2}/L}\sum_{j \neq
    i}\left[\sin(k_j x_j+\phi^{(k)}_j(t))\right],
\label{eq:turbo}
\end{equation}
where $i,j=1,2,3$, $A$ is the forcing amplitude, $\mathbf{k}$ is the
wavevector, and the sum is limited to
$k^{2}=k_{1}^{2}+k_{2}^{2}+k_{3}^{2}\leq 8 \pi^2/L^2$.  The phases
$\phi_j^{(k)}$ are evolved in time according to independent
Ornstein-Uhlenbeck processes with the same relaxation times
$T_L=L/U_{\text{rms}}$, where $L$ is the cubic box edge and
$U_{\text{rms}}$ is a typical large-scale
velocity~\citep{Biferale_2011,perlekar2012droplet}. All the data
presented in this work come from simulations with $L=512$.

\subsection{The volume fraction correction in numerical simulations}\label{subsec:vol-corr}

\begin{figure}[htbp]
\centering
\includegraphics[width=0.8\textwidth]{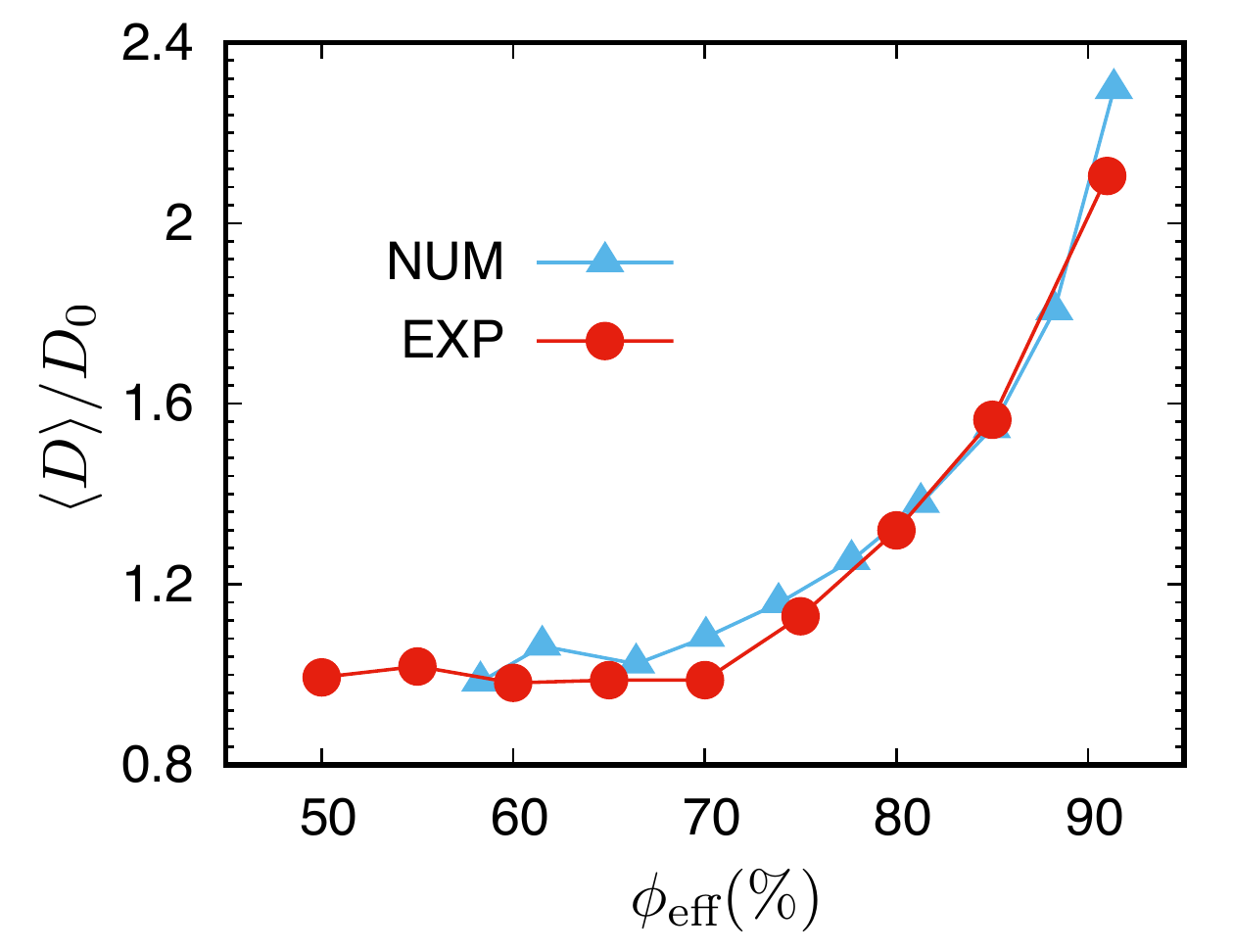}
\caption{ Mean droplet diameter, $\langle D \rangle$ (normalized by
  $D_0$), as a function of the $\phi$ from experiments and numerical
  simulations $\phi_{eff}$ (multiplied by the scaling factor
  $(1+3h/\langle D \rangle)$, where $h \approx 2.5$ is the interface
  width).  }
\label{figure:Dmean}
\end{figure}

Beyond a given volume fraction of the droplet phase immersed in a
continuous fluid film, $\phi=55-65\%$, packed emulsion in a dynamic
population regime can turn into CPI when reaching the critical volume
fraction, $\phi_c$. This is characterized by several factors such as,
for instance, the fluid components (i.e., fluid density), the effect
of the surfactant and the hydrodynamic stirring. Even more factors
come into play when comparing numerical simulations with experiments
where also the length and time scales differ. To provide a closer
approximation between the computation of $\phi$ in experiments and
numerical simulations we introduce a factor $h$, that characterizes
the representation of the interface width between the two fluid
components in numerical simulation, and that helps to provide a better
estimation of $\phi$ with respect to experiments. In
Fig.~\ref{figure:Dmean} we report an estimation of the mean droplet
diameter, $\langle D \rangle$, at the growing of $\phi$ comparing a
typical process of emulsification via numerical simulation and
experiments. We then define $\phi_{eff}$ to indicate the critical
volume fraction for a given set of a parameters that lead a numerical
simulation to CPI.

%%%%%%%%%%%%%%%%%%%%%%%%
%%%%%%%%%%%%%%%%%%%%%%%%%%

\subsection{Population balance models}\label{subsec:pbm}
\begin{figure}[htbp]
\centering
\includegraphics[width=1\textwidth]{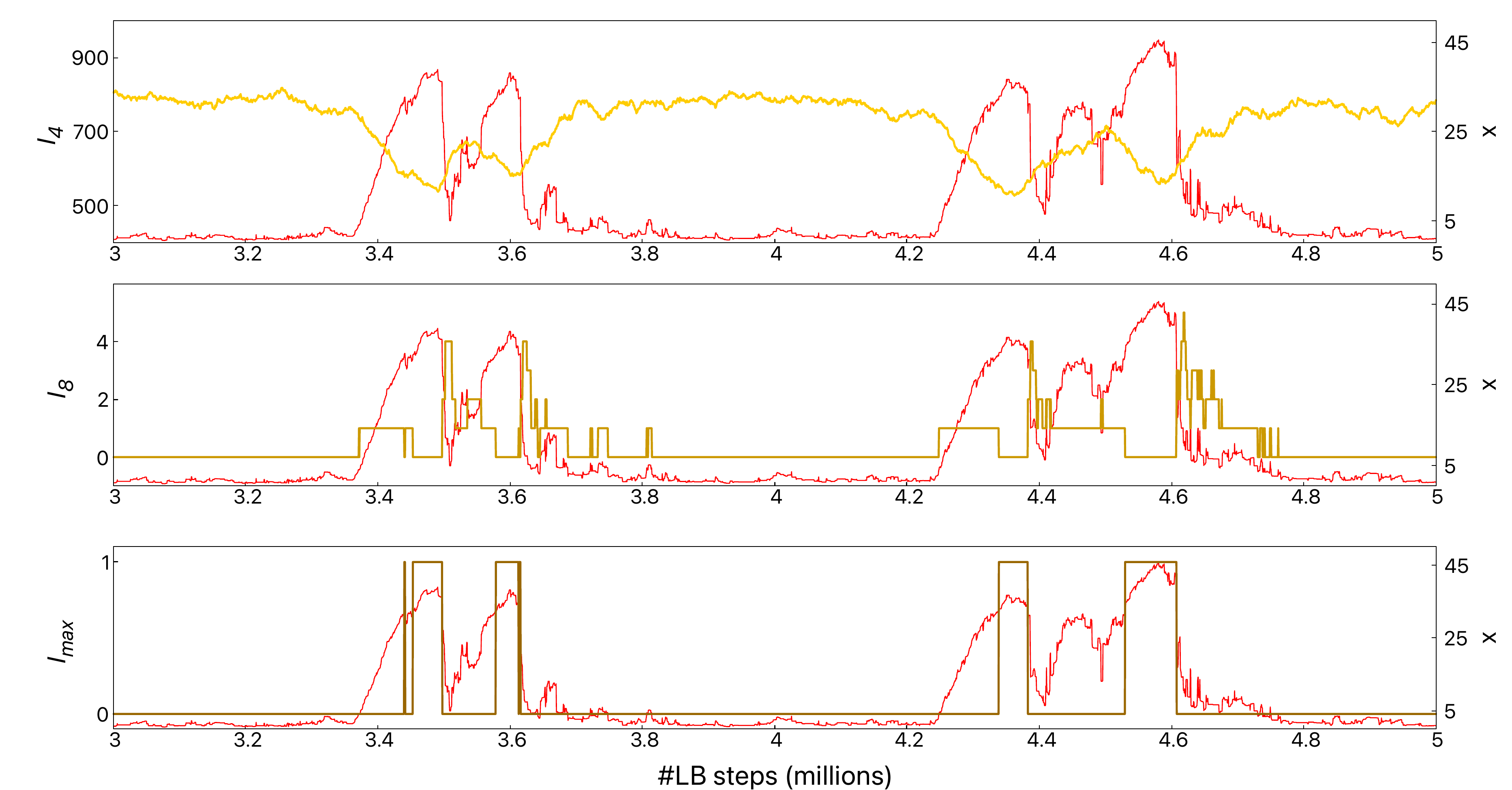}
\caption{A graphical visualization of the temporal evolution of the
  cumulative distribution $\overline{n}_I(t) = \sum_{i \in I} n_i(t)$
  for a few representative size-classes of droplets during a typical
  emulsification process via fully resolved numerical simulations:
  from small size ($I_{4} = \{i| R_i \leq 4\langle R \rangle \}$), to
  large size ($I_8 = \{i| 4 \langle R\rangle <R_i \leq 8\langle R
  \rangle \}$) and up to the largest-size ($I_{\text{max}} = \{i| R_i
  = R_{\text{max}}\}$), where $R_i$ is the radius of the $i$-th
  droplet and $\langle R \rangle$ is the mean radius over the ensemble
  of droplets. The picture displays the population dynamics (y,
  different shades of yellow) overlapped with the time evolution of
  the fraction of volume occupied by the largest droplet $x(t)$ (y2,
  red). It provides evidence of a slower coalescence-driven dynamics
  coinciding with the formation of a large droplet, followed by a
  faster breakup-driven dynamics that tends to bring the emulsion back
  to equilibrium. It shows that during the coalescence-driven phase
  only a single large droplet is formed in the emulsion that becomes
  bigger and bigger by including small droplets. Indeed, the number of
  droplets of the small-size class generally decreases while only one
  single increasing large droplet is displayed on the plots related to
  the larger-size classes. On the other hand, the largest droplets
  breaks up forming multiple large droplets of the same size-class,
  with this behavior repeating recursively on the newly generated
  droplets until equilibrium, whether not leading to the formation of
  a new large droplet region or eventually at a CPI.}
\label{figure:popdyn_2}
\end{figure}

To better highlight the morphological changes in the emulsion which
become essential for the dynamics at approaching the critical volume
fraction, we focus on a population description of the emulsion. In
this approach, we disregard the space dimensionality and the
hydrodynamics and elastic processes between droplets, to focus only on
the dynamics of droplet sizes. The key description of the systems,
which is considered effectively zero-dimensional, is the distribution
of number of droplets $n_i(t) \equiv n(x_i,t)$ per size class, $x_i$,
as a function of time. The size classes can be taken as geometrically
distributed multiples of some characteristic droplet size (e.g. the
Kolmogorov-Hinze size), or submultiple thereof, i.e. $x_i=x_{KH}
\lambda^{i}$.  In the static population dynamics regime (see Figure
\ref{figure:pop_dynamics}) the number of droplets in each size class
does change in the initial transient phase, when the emulsion is
formed or when some parameter is adjusted, and then does not depend on
time anymore. On the contrary, in the population dynamic regime (see
Figure \ref{figure:pop_dynamics}) the droplet distribution is
constantly changing over time and fluctuating around some average
value (see Figure \ref{figure:pop_dynamics}). Such type of processes
are typically described in terms of a population balance equation
(PBE), whose most generic form reads \cite{Coulaloglou1976,Maass2012}:
\begin{equation}\label{eq:pbe}
  \dot{n}_i = B^{(b)}_i - D^{(b)}_i + B^{(c)}_i - D^{(c)}_i\quad \mbox{for} \; i=1,2,\dots,M
\end{equation}  
where $B_i^{(b,c)}$ and $D_i^{(b,c)}$ are birth and death rates of
droplets in the $i$-th class due to breakup and coalescence,
respectively.  A more explicit instance of a PBE can be written as the
following Smoluchowski-like aggregation-fragmentation equation
\cite{Krapivsky2010,Brilliantov2015}:
\begin{equation}\label{eq:smolu}
  \dot{n}_i = \sum_{j=1}^{M-1} p_j \beta_{i+j} n_{i+j} + p_i \sum_{j=i+1}^{M} \beta_j n_j
  - \left(\sum_{j=1}^{i-1} p_j \right) \beta_i n_i
    -\sum_{j=1}^{M-i}\kappa_{i,j}n_j n_i + \frac{1}{2}\sum_{j=1}^{i-1}\kappa_{j,i-j}n_jn_{i-j} 
\end{equation}  
The above equation provides a description of the time-dependent change
of $n_i(t)$ due to the coalescence (with rates $\kappa_{i,j}$) and
breakup (with rates $\beta_i$).  This Smoluchowski
aggregation-fragmentation equation describes the change in sizes of
droplets and the rates, $\kappa_{i,j}$ and $\beta_i$, are functions of
the underlying fluid dynamic processes that control the coalescence
and breakup events; these include the Reynolds number of the flow, the
surface tension, and disjoining pressure, as well as the volume
fraction.  From the fully resolved numerical simulation, the rates
$\kappa_{i,j}$ and $\beta_i$ can be accurately measured. A simplified
representation of population dynamics as emerging from the simulations
is reported in Figure \ref{figure:popdyn_2}.

\subsection{Non reciprocal dynamical model}\label{subsec:model}
In this section we further simplify the population model based on the
Smoluchowski equation into a simpler model already discussed in the
main part of the paper. The model aims at capturing the essential
features of the dynamics in proximity of the critical volume
fraction. The basic starting point is the consideration of the fact
that at approaching the critical volume fraction the emulsion starts
to develop larger regions of partially inverted emulsion. These
regions, visible in Figure \ref{figure:popdyn_2}, can be represented
in terms of the volume of the largest droplets. We therefore assume
that at approaching the critical volume fraction the Smoluchowski
dynamics is actually controlled by the presence of a very large number
of small droplets of size of the order of the Hinze radius, $R_H$ and
by one large droplet that, as a function of time, can grow, due to
coalescence with smaller droplets, and shrink, due to breakup into
smaller droplets. In these conditions the two dynamical variables
essential to describe the dynamics are then the volume of the largest
droplet, $x$, and the number of small droplets.  The key physical
feature of the dynamics is the presence of two physically distinct
phenomena, one associated to droplets coalescence and the other with
droplets breakup. The typical timescales of the two processes are
different, as well as their dependency on the underlying hydrodynamic
processes.  Next, we assume that, close to the phase inversion, the
statistical property of the emulsion can be described in terms of the
largest droplet volume $V_{\text{max}}$. Then, phase inversion occurs
when $V_{\text{max}}$ reaches its maximum available size $V_M$. By
defining $V_H$ the minimum droplet size for $V_{\text{max}}$, we
introduce the dimensionless variable:
\begin{equation}
    \label{m1}
    x \equiv \frac{V_{\text{max}}-V_H}{V_M-V_H}
\end{equation}
The next step is to model the dynamical behaviour of $x \in [0,1]$
subject to breakup and coalescence. Let $\phi$ the volume
fraction. For $\phi > 50\%$, emulsions are subject to coarsening whose
dynamics depends on $\phi$, the effects of surfactants (if present)
and turbulence. Eventually, for $x=1$ coalescence process
stops. Hereafter, we assume that the above dynamics is described by
the equation:
\begin{equation}
    \label{m2}
    \frac{dx}{dt} = \alpha \phi (1-x)
\end{equation}
with $\alpha \ge 0$. Next, let us consider the process of breaking up,
which counteracts the dynamics of coalescence. This can be done by
writing:
\begin{equation}
    \label{m3}
    \frac{dx}{dt} = (\alpha \phi -fx)  (1-x)
\end{equation}
where $f \ge 0$. The above equation is the simplest form of
coalescence and breakup dynamics of $x$. Now, we need to specify the
two unknown quantities $\alpha$ and $f$. From eq. (\ref{m3}) we see
that, for $\alpha \phi/f < 1$, the fixed point $x_*=\alpha \phi/ f$ is
stable. Inspection of the inset of figure \ref{figure:popdyn_2} tells
that $x$ is a rather intermittent quantity showing relatively long
periods with $x\sim 0$ and bursts of large value of $x$ close to
$1$. This observation suggests that the ratio $\alpha/f$ should be an
intermittent quantity which we need to model. To do that, we assume
$\alpha$ to be constant and $f$ to satisfy the stochastic differential
equation:
\begin{equation}
    \label{m4}
    df = \left[-r f - \beta f^3 + \frac{1}{4} \frac{d\xi}{df}\right] + \sqrt{\xi(f,x)} dW(t).
\end{equation}
A reasonable physical ground for this equation is provided in what
follows. The first two terms in the square bracket represent the
linear and the non linear decreases of $f$ towards the state $f=0$
where (for zero noise) the system does not exhibit breakup
events. This situation should occurs when $x$ is close to $1$.  The
term $\xi(f,x)$ is the variance of the noise introduced in
eq. (\ref{m4}) to mimic, in a very simple way, the effect of
turbulence. Then the last term in the square brackets in
eq. (\ref{m4}) is the Stratonovich term induced by multiplicative
noise \cite{Risken}. For $x$ relatively small, we expect turbulence to
destabilize the state $f=0$. This implies that $\xi(f,x)$ should be at
least quadratic in $f$. Also we expect that this effect disappear for
$x$ close to $1$.  The above discussion suggests the following
expression for $\xi(f,x)$:
\begin{equation}
    \label{m5}
    \xi(f,x) = \varepsilon_0 + \varepsilon_1 f^2 (1-x)
\end{equation}
Eq. (\ref{m4}) now reads:
\begin{equation}
    \label{m6}
   df = [-r f -\beta f^3 + \frac{1}{2} \varepsilon_1 f(1-x)]dt + \sqrt{\varepsilon_0+\varepsilon_1f^2(1-x)} dW(t)
\end{equation}
Notice that assuming $x$ constant, the stationary probability distribution $P(f|x)$ is
\begin{equation}
    \label{m7}    
     P(f) =\frac{Z}{[\varepsilon_0+\varepsilon_1 f^2(1-x)]^{1/2+r/(\varepsilon_1(1-x))}}\exp(-N_L)
\end{equation}
where $Z$ is a normalization constant and $N_L \sim f^2 + ...$ stands
for the contributions of the non linear terms.  From eq. (\ref{m7} it
is clear that, for $x<1$, $P(f)$ shows a power law distribution at
relatively large $f$ which is the signature of intermittency. Notice
that eq.(\ref{m5}) and eq. (\ref{m6}) are meaningful for $f\ge 0$,
i.e. $\partial_f P(f)|_{f=0}=0$. \\ First of all, let us discuss the
case $\varepsilon_1=0$. In this case, $f$ is spending most of the time
near $f=0$ with a variance $f_s \sim \sqrt{\varepsilon_0/(2r)}$. Then
the value of $x$ stabilizes near the value $ \alpha \phi/f_s$.
Therefore, phase inversion occurs at $\phi_s = f_s/\alpha$. It is
simple to check that this is also the critical value of $\phi$ at
which the phase inverted state $x=1$ is stable for any value of
$\varepsilon_1$. Thus $\phi_s$ should be considered the lower bound at
which phase inversion can occur. Let us remark that, for
$\varepsilon_1=0$, the system exhibits one and only one stable state
$\alpha \phi /f_s$ for $\phi \le \phi_s$ and $x=1 $ for $\phi \ge
\phi_s$. In other words, no sharp transition (catastrophic phase
inversion) is observed and the system linearly increase the average
value of $x$ up to $x=1$

\begin{figure}[htbp]
\centering
\includegraphics[width=0.7\textwidth]{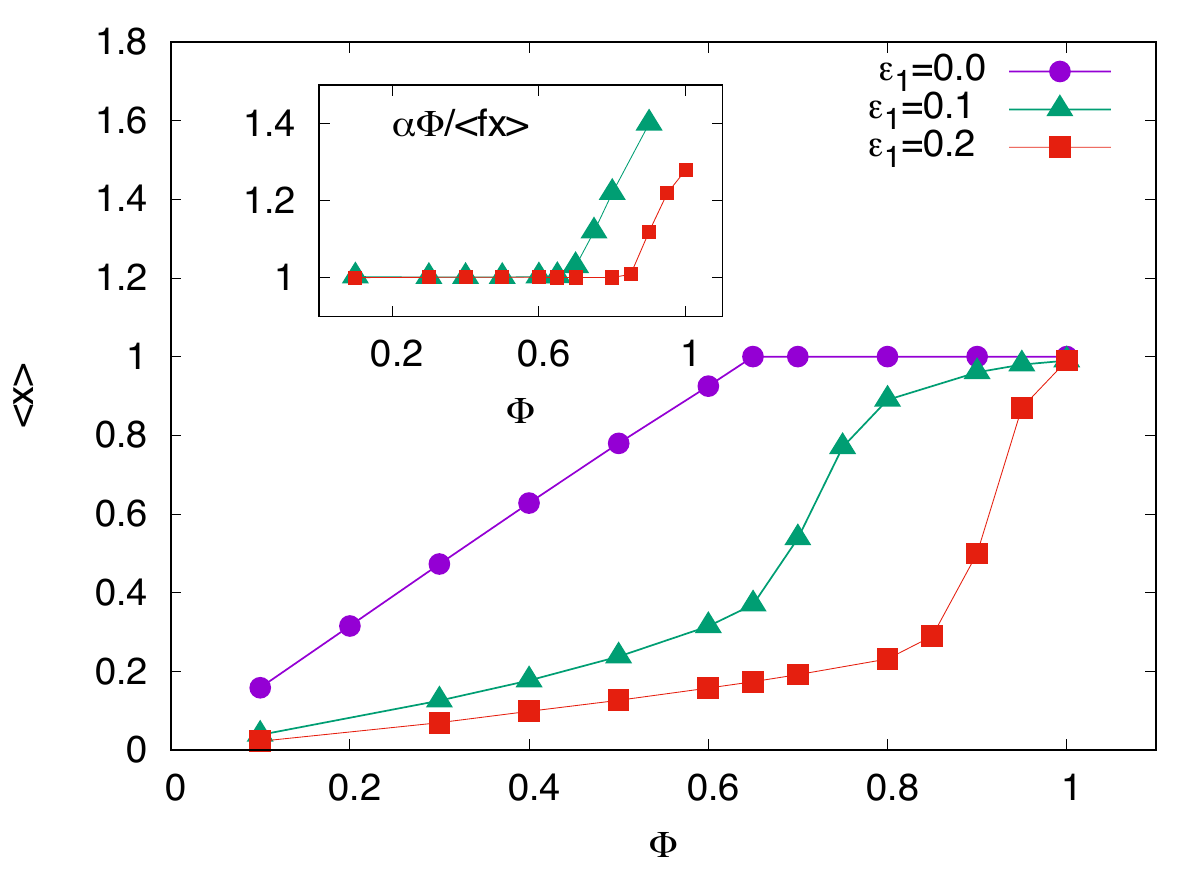}
\caption{Main panel. Behavior of the average value $\langle x \rangle$
  for three values of $\varepsilon_1=0$ (circles), $0.1$ (triangles)
  and $0.2$ (squares) and for fixed $\varepsilon_0=10^{-7}$.  Notice
  that for increasing value of $\varepsilon_1$ the transition to
  $\langle x \rangle =1$ occurs for larger values of $\phi$. Each
  point in the figure is obtained by averaging in time and over 100
  independent realizations.  Inset: The ratio of $\alpha \phi /
  \langle f x \rangle $ as a function of $\phi$ for $\varepsilon_1 =
  0.1, 0.2$ (same symbols as in the main panel). The results clearly
  show that for small enough $\phi$ (depending on $\varepsilon_1$) the
  system shows stationary states for $\langle f x\rangle = \alpha
  \phi$. }
\label{figure:model_behavior}
\end{figure}

The situation changes completely for the non-vanishing value of
$\varepsilon_1$.  From eq. (\ref{m6}) we can compute the critical
value $x=x_s$ at which the linear term in square bracket of
eq. (\ref{m6}) becomes zero. A simple computation gives
$x_s=1-2r/\varepsilon_1$. This result can be interpreted by saying
that a phase inversion can occur for $\phi \ge \phi_s$ only if $x$
becomes larger than $x_s$. Notice that when $x\ge x_s$ with
$\varepsilon_1>0$ the system is characterized by two different time
scales: the coalescence time scale $1/(\alpha \phi)$ and the breakup
time scale associated to the tail of the probability distribution
$P(f)$, i.e. to the characteristic time for $f$ to show an {\it
  intermittent excursion}. The above discussion implies that the
transition to the (stable) state $x=1$ occurs at a critical value
$\phi=\phi_c$ with $\phi_c > \phi_s$. Also $\phi_c$ increases with
$\varepsilon_1$. For $\phi <\phi_c$ the system is characterized by the
balance of $\alpha \phi$ with $\langle f x \rangle$, where
$\langle.. \rangle$ stands for time average, whereas for $\phi >
\phi_c$ one has $\alpha \phi > \langle f x \rangle$. Technically this
also implies that near the transition there exist two stable states,
namely the phase inverted state $x=1$ and the jammed state $\alpha
\phi = \langle f x \rangle$.  \\ To validate the above discussion, we
consider the following model parameters: $\alpha=0.002$, $r=0.02$,
$\varepsilon_0=10^{-7}$ and two choices of $\varepsilon_1$ namely
$\varepsilon_1 =0.1$ and $\varepsilon_1 = 0.2$.  The value of $\alpha$
is chosen by tuning our model with the LBM simulations. Numerical
simulations for $\varepsilon_1=0$ gives the estimate $\phi_s =
0.65$. In figure~\ref{figure:model_behavior} we show the average value
of $x$, denoted by $\langle x \rangle$ a function of $\phi$ obtained
by averaging over $100$ noise realizations. In the main part of the
figure we plot $\langle x \rangle$ for $\phi$ increasing from $0.1$
for $\varepsilon_1=0$ (circles), $0.1$ (triangles) and $0.2$
(squares).  Upon increasing $\varepsilon_1$ we observe a shift in the
value of $\phi_c$ which increases with $\varepsilon_1$. In the insert
we show the quantity $Q\equiv \alpha \phi/\langle fx \rangle$. For
$\phi\le \phi_c$, $Q$, this ratio is close to $1$, i.e. the system
shows balance between the coalescence terem $\alpha \phi$ and the
fragmentation $\langle fx \rangle$. Close to $\phi=\phi_c$, $Q$ starts
to grow while the average value of $x$ increases towards the phase
inverted state $x=1$. %Finally, in figure (\ref{figure:model_sim}) we
compare a snapshot of the time behavior of\\ The theoretical
discussion and the results shown in figure \ref{figure:model_behavior}
indicate that the system does exhibit a sharp transition between the
jammed state (small $\langle x \rangle$) and a phase inverted state
$x=1$, i.e. a catastrophic phase inversion.

Since the model depends on many parameters, it is worthwhile to show
that it can be recast in a somehow simpler way. Using the quantity
$f_s$, $\phi_s$ and $x_s$, upon rescaling the time as $\tilde t = r t
$, the rate of fragmentation $\tilde f = f / f_s$, it is possible to
rewrite the model in the following way:
\begin{eqnarray}
    \label{m8}
    d\tilde f &=& \left[ \frac{x_s-x}{1-x_s}\tilde f- \tilde \beta \tilde f^3\right] d\tilde t + \sqrt{2+2\tilde f^2 \frac{1-x}{1-x_s}} dW(\tilde t) \\
    \label{m9}
    \frac{dx}{d\tilde t} &=& A\left[ \frac{\phi}{\phi_s}-\tilde f x\right](1-x)
\end{eqnarray}
where $A=f_s/r$ and $\tilde\beta = f_s^2 \beta/r$ are dimensionless
parameters.  The above equations show that there are three independent
parameters namely $\phi_s$, $x_s$ and $A$. The latter can be written
as $A = \phi_s \alpha/r$ and it is proportional to the ratio of two
time scales characterizing the dynamics of the system, i.e. the
coalescence and break up processes. Based on the numerical simulations
previously discussed (see also \cite{yi2024divergence}) we can assume
that coalescence time scale $1/\alpha$ is much longer than the breakup
time scale $1/r$ which implies $A\ll 1$ Once the value of $A$ is
fixed, the model depends on the two parameter $\phi_s$ (the lower
bound for the phase inversion to occur) and $x_c$ (the value of
$V_{max}$ above which the system may be driven by the noise to a phase
inversion). For the numerical results previously discussed, the
corresponding value of $A$ in eq. (\ref{m9}) is $A=0.08$.

\bigskip

\section*{Acknowledgments}
  This work is financially supported by the National Natural Science
  Foundation of China (No. 12588201), the Netherlands Organisation for
  Scientific Research (NWO) for the use of supercomputer facilities
  (Snellius) under Grant No. 2021.035, 2023.026, and the New
  Cornerstone Science Foundation through the New Cornerstone
  Investigator Program and the XPLORER PRIZE. Numerical simulations
  were performed thanks to granted PRACE Projects (IDs No. 2018184340
  and No. 2019204899) along with CINECA and BSC for access to their
  HPC systems.

\bibliography{main}

\begin{thebibliography}{41}%
\makeatletter
\providecommand \@ifxundefined [1]{%
 \@ifx{#1\undefined}
}%
\providecommand \@ifnum [1]{%
 \ifnum #1\expandafter \@firstoftwo
 \else \expandafter \@secondoftwo
 \fi
}%
\providecommand \@ifx [1]{%
 \ifx #1\expandafter \@firstoftwo
 \else \expandafter \@secondoftwo
 \fi
}%
\providecommand \natexlab [1]{#1}%
\providecommand \enquote  [1]{``#1''}%
\providecommand \bibnamefont  [1]{#1}%
\providecommand \bibfnamefont [1]{#1}%
\providecommand \citenamefont [1]{#1}%
\providecommand \href@noop [0]{\@secondoftwo}%
\providecommand \href [0]{\begingroup \@sanitize@url \@href}%
\providecommand \@href[1]{\@@startlink{#1}\@@href}%
\providecommand \@@href[1]{\endgroup#1\@@endlink}%
\providecommand \@sanitize@url [0]{\catcode `\\12\catcode `\$12\catcode
  `\&12\catcode `\#12\catcode `\^12\catcode `\_12\catcode `\%12\relax}%
\providecommand \@@startlink[1]{}%
\providecommand \@@endlink[0]{}%
\providecommand \url  [0]{\begingroup\@sanitize@url \@url }%
\providecommand \@url [1]{\endgroup\@href {#1}{\urlprefix }}%
\providecommand \urlprefix  [0]{URL }%
\providecommand \Eprint [0]{\href }%
\providecommand \doibase [0]{https://doi.org/}%
\providecommand \selectlanguage [0]{\@gobble}%
\providecommand \bibinfo  [0]{\@secondoftwo}%
\providecommand \bibfield  [0]{\@secondoftwo}%
\providecommand \translation [1]{[#1]}%
\providecommand \BibitemOpen [0]{}%
\providecommand \bibitemStop [0]{}%
\providecommand \bibitemNoStop [0]{.\EOS\space}%
\providecommand \EOS [0]{\spacefactor3000\relax}%
\providecommand \BibitemShut  [1]{\csname bibitem#1\endcsname}%
\let\auto@bib@innerbib\@empty
%</preamble>
\bibitem [{\citenamefont {Rouwhorst}\ \emph {et~al.}(2020)\citenamefont
  {Rouwhorst}, \citenamefont {Ness}, \citenamefont {Stoyanov}, \citenamefont
  {Zaccone},\ and\ \citenamefont {Schall}}]{Rouwhorst2020}%
  \BibitemOpen
  \bibfield  {author} {\bibinfo {author} {\bibfnamefont {J.}~\bibnamefont
  {Rouwhorst}}, \bibinfo {author} {\bibfnamefont {C.}~\bibnamefont {Ness}},
  \bibinfo {author} {\bibfnamefont {S.}~\bibnamefont {Stoyanov}}, \bibinfo
  {author} {\bibfnamefont {A.}~\bibnamefont {Zaccone}},\ and\ \bibinfo {author}
  {\bibfnamefont {P.}~\bibnamefont {Schall}},\ }\bibfield  {title} {\bibinfo
  {title} {Nonequilibrium continuous phase transition in colloidal gelation
  with short-range attraction},\ }\href@noop {} {\bibfield  {journal} {\bibinfo
   {journal} {Nat. Commun.}\ }\textbf {\bibinfo {volume} {11}} (\bibinfo {year}
  {2020})}\BibitemShut {NoStop}%
\bibitem [{\citenamefont {Paradossi}\ \emph {et~al.}(2002)\citenamefont
  {Paradossi}, \citenamefont {Chiessi}, \citenamefont {Barbiroli},\ and\
  \citenamefont {Fessas}}]{Paradossi2002}%
  \BibitemOpen
  \bibfield  {author} {\bibinfo {author} {\bibfnamefont {G.}~\bibnamefont
  {Paradossi}}, \bibinfo {author} {\bibfnamefont {E.}~\bibnamefont {Chiessi}},
  \bibinfo {author} {\bibfnamefont {A.}~\bibnamefont {Barbiroli}},\ and\
  \bibinfo {author} {\bibfnamefont {D.}~\bibnamefont {Fessas}},\ }\bibfield
  {title} {\bibinfo {title} {Xanthan and glucomannan mixtures: synergistic
  interactions and gelation},\ }\href@noop {} {\bibfield  {journal} {\bibinfo
  {journal} {Biomacromolecules}\ }\textbf {\bibinfo {volume} {3}},\ \bibinfo
  {pages} {498} (\bibinfo {year} {2002})}\BibitemShut {NoStop}%
\bibitem [{\citenamefont {Li}\ \emph {et~al.}(2025)\citenamefont {Li},
  \citenamefont {Salmon}, \citenamefont {Hassaini}, \citenamefont {Chang},
  \citenamefont {Mucignat},\ and\ \citenamefont {Coletti}}]{Li2025}%
  \BibitemOpen
  \bibfield  {author} {\bibinfo {author} {\bibfnamefont {Y.}~\bibnamefont
  {Li}}, \bibinfo {author} {\bibfnamefont {H.}~\bibnamefont {Salmon}}, \bibinfo
  {author} {\bibfnamefont {R.}~\bibnamefont {Hassaini}}, \bibinfo {author}
  {\bibfnamefont {K.}~\bibnamefont {Chang}}, \bibinfo {author} {\bibfnamefont
  {C.}~\bibnamefont {Mucignat}},\ and\ \bibinfo {author} {\bibfnamefont
  {F.}~\bibnamefont {Coletti}},\ }\bibfield  {title} {\bibinfo {title}
  {Spatiotemporal scales of motion and particle clustering in free-surface
  turbulence},\ }\href@noop {} {\bibfield  {journal} {\bibinfo  {journal}
  {Phys. Rev. Fluids}\ }\textbf {\bibinfo {volume} {10}} (\bibinfo {year}
  {2025})}\BibitemShut {NoStop}%
\bibitem [{\citenamefont {C\'ozar}\ \emph {et~al.}(2014)\citenamefont
  {C\'ozar}, \citenamefont {Echevarr\'{\i}a}, \citenamefont
  {Gonz\'alez-Gordillo}, \citenamefont {Irigoien}, \citenamefont {\'Ubeda},
  \citenamefont {Hern\'andez-Le\'on}, \citenamefont {Palma}, \citenamefont
  {Navarro}, \citenamefont {Garc\'{\i}a-de Lomas}, \citenamefont {Ruiz},
  \citenamefont {Fern\'andez-de Puelles},\ and\ \citenamefont
  {Duarte}}]{Cozar2014}%
  \BibitemOpen
  \bibfield  {author} {\bibinfo {author} {\bibfnamefont {A.}~\bibnamefont
  {C\'ozar}}, \bibinfo {author} {\bibfnamefont {F.}~\bibnamefont
  {Echevarr\'{\i}a}}, \bibinfo {author} {\bibfnamefont {J.}~\bibnamefont
  {Gonz\'alez-Gordillo}}, \bibinfo {author} {\bibfnamefont {X.}~\bibnamefont
  {Irigoien}}, \bibinfo {author} {\bibfnamefont {B.}~\bibnamefont {\'Ubeda}},
  \bibinfo {author} {\bibfnamefont {S.}~\bibnamefont {Hern\'andez-Le\'on}},
  \bibinfo {author} {\bibfnamefont {A.}~\bibnamefont {Palma}}, \bibinfo
  {author} {\bibfnamefont {S.}~\bibnamefont {Navarro}}, \bibinfo {author}
  {\bibfnamefont {J.}~\bibnamefont {Garc\'{\i}a-de Lomas}}, \bibinfo {author}
  {\bibfnamefont {A.}~\bibnamefont {Ruiz}}, \bibinfo {author} {\bibfnamefont
  {M.}~\bibnamefont {Fern\'andez-de Puelles}},\ and\ \bibinfo {author}
  {\bibfnamefont {C.}~\bibnamefont {Duarte}},\ }\bibfield  {title} {\bibinfo
  {title} {Plastic debris in the open ocean},\ }\href@noop {} {\bibfield
  {journal} {\bibinfo  {journal} {Proc. Natl. Acad. Sci. USA}\ }\textbf
  {\bibinfo {volume} {111}},\ \bibinfo {pages} {10239} (\bibinfo {year}
  {2014})}\BibitemShut {NoStop}%
\bibitem [{\citenamefont {Wang}\ \emph {et~al.}(2021)\citenamefont {Wang},
  \citenamefont {Bolan}, \citenamefont {Tsang}, \citenamefont {Sarkar},
  \citenamefont {Bradney},\ and\ \citenamefont {Li}}]{Wang2021}%
  \BibitemOpen
  \bibfield  {author} {\bibinfo {author} {\bibfnamefont {X.}~\bibnamefont
  {Wang}}, \bibinfo {author} {\bibfnamefont {N.}~\bibnamefont {Bolan}},
  \bibinfo {author} {\bibfnamefont {D.}~\bibnamefont {Tsang}}, \bibinfo
  {author} {\bibfnamefont {B.}~\bibnamefont {Sarkar}}, \bibinfo {author}
  {\bibfnamefont {L.}~\bibnamefont {Bradney}},\ and\ \bibinfo {author}
  {\bibfnamefont {Y.}~\bibnamefont {Li}},\ }\bibfield  {title} {\bibinfo
  {title} {A review of microplastics aggregation in aquatic environment:
  Influence factors, analytical methods, and environmental implications},\
  }\href@noop {} {\bibfield  {journal} {\bibinfo  {journal} {J. Hazard.
  Mater.}\ }\textbf {\bibinfo {volume} {402}} (\bibinfo {year}
  {2021})}\BibitemShut {NoStop}%
\bibitem [{\citenamefont {Morbidelli}\ \emph {et~al.}(2012)\citenamefont
  {Morbidelli}, \citenamefont {Lunine}, \citenamefont {O'Brien}, \citenamefont
  {Raymond},\ and\ \citenamefont {Walsh}}]{Morbidelli2012}%
  \BibitemOpen
  \bibfield  {author} {\bibinfo {author} {\bibfnamefont {A.}~\bibnamefont
  {Morbidelli}}, \bibinfo {author} {\bibfnamefont {J.}~\bibnamefont {Lunine}},
  \bibinfo {author} {\bibfnamefont {D.}~\bibnamefont {O'Brien}}, \bibinfo
  {author} {\bibfnamefont {S.}~\bibnamefont {Raymond}},\ and\ \bibinfo {author}
  {\bibfnamefont {K.}~\bibnamefont {Walsh}},\ }\bibfield  {title} {\bibinfo
  {title} {Building terrestrial planets},\ }\href@noop {} {\bibfield  {journal}
  {\bibinfo  {journal} {Annu. Rev. Earth Planet. Sci.}\ }\textbf {\bibinfo
  {volume} {40}},\ \bibinfo {pages} {251} (\bibinfo {year} {2012})}\BibitemShut
  {NoStop}%
\bibitem [{\citenamefont {Yanagisawa}\ \emph {et~al.}(2021)\citenamefont
  {Yanagisawa}, \citenamefont {Tani},\ and\ \citenamefont
  {Kurita}}]{yanagisawa2021dynamics}%
  \BibitemOpen
  \bibfield  {author} {\bibinfo {author} {\bibfnamefont {N.}~\bibnamefont
  {Yanagisawa}}, \bibinfo {author} {\bibfnamefont {M.}~\bibnamefont {Tani}},\
  and\ \bibinfo {author} {\bibfnamefont {R.}~\bibnamefont {Kurita}},\
  }\bibfield  {title} {\bibinfo {title} {Dynamics and mechanism of liquid film
  collapse in a foam},\ }\href@noop {} {\bibfield  {journal} {\bibinfo
  {journal} {Soft Matter}\ }\textbf {\bibinfo {volume} {17}},\ \bibinfo {pages}
  {1738} (\bibinfo {year} {2021})}\BibitemShut {NoStop}%
\bibitem [{\citenamefont {Wang}\ \emph {et~al.}(2016)\citenamefont {Wang},
  \citenamefont {Nguyen},\ and\ \citenamefont {Farrokhpay}}]{Wang201655}%
  \BibitemOpen
  \bibfield  {author} {\bibinfo {author} {\bibfnamefont {J.}~\bibnamefont
  {Wang}}, \bibinfo {author} {\bibfnamefont {A.~V.}\ \bibnamefont {Nguyen}},\
  and\ \bibinfo {author} {\bibfnamefont {S.}~\bibnamefont {Farrokhpay}},\
  }\bibfield  {title} {\bibinfo {title} {A critical review of the growth,
  drainage and collapse of foams},\ }\href@noop {} {\bibfield  {journal}
  {\bibinfo  {journal} {Adv. Colloid Interface Sci.}\ }\textbf {\bibinfo
  {volume} {228}},\ \bibinfo {pages} {55} (\bibinfo {year} {2016})}\BibitemShut
  {NoStop}%
\bibitem [{\citenamefont {Perazzo}\ \emph {et~al.}(2015)\citenamefont
  {Perazzo}, \citenamefont {Preziosi},\ and\ \citenamefont
  {Guido}}]{perazzo2015phase}%
  \BibitemOpen
  \bibfield  {author} {\bibinfo {author} {\bibfnamefont {A.}~\bibnamefont
  {Perazzo}}, \bibinfo {author} {\bibfnamefont {V.}~\bibnamefont {Preziosi}},\
  and\ \bibinfo {author} {\bibfnamefont {S.}~\bibnamefont {Guido}},\ }\bibfield
   {title} {\bibinfo {title} {Phase inversion emulsification: Current
  understanding and applications},\ }\href@noop {} {\bibfield  {journal}
  {\bibinfo  {journal} {Adv. Colloid Interface Sci.}\ }\textbf {\bibinfo
  {volume} {222}},\ \bibinfo {pages} {581} (\bibinfo {year}
  {2015})}\BibitemShut {NoStop}%
\bibitem [{\citenamefont {Kumar}\ \emph {et~al.}(2015)\citenamefont {Kumar},
  \citenamefont {Li}, \citenamefont {Cheng},\ and\ \citenamefont
  {Lee}}]{kumar2015recent}%
  \BibitemOpen
  \bibfield  {author} {\bibinfo {author} {\bibfnamefont {A.}~\bibnamefont
  {Kumar}}, \bibinfo {author} {\bibfnamefont {S.}~\bibnamefont {Li}}, \bibinfo
  {author} {\bibfnamefont {C.-M.}\ \bibnamefont {Cheng}},\ and\ \bibinfo
  {author} {\bibfnamefont {D.}~\bibnamefont {Lee}},\ }\bibfield  {title}
  {\bibinfo {title} {Recent developments in phase inversion emulsification},\
  }\href@noop {} {\bibfield  {journal} {\bibinfo  {journal} {Ind. Eng. Chem.
  Res.}\ }\textbf {\bibinfo {volume} {54}},\ \bibinfo {pages} {8375} (\bibinfo
  {year} {2015})}\BibitemShut {NoStop}%
\bibitem [{\citenamefont {Leal-Calderon}\ \emph {et~al.}(2007)\citenamefont
  {Leal-Calderon}, \citenamefont {Schmitt},\ and\ \citenamefont
  {Bibette}}]{leal2007emulsion}%
  \BibitemOpen
  \bibfield  {author} {\bibinfo {author} {\bibfnamefont {F.}~\bibnamefont
  {Leal-Calderon}}, \bibinfo {author} {\bibfnamefont {V.}~\bibnamefont
  {Schmitt}},\ and\ \bibinfo {author} {\bibfnamefont {J.}~\bibnamefont
  {Bibette}},\ }\href@noop {} {\emph {\bibinfo {title} {Emulsion science: basic
  principles}}}\ (\bibinfo  {publisher} {Springer Science \& Business Media},\
  \bibinfo {year} {2007})\BibitemShut {NoStop}%
\bibitem [{\citenamefont {Vaessen}\ and\ \citenamefont
  {Stein}(1995)}]{Vaessen1995}%
  \BibitemOpen
  \bibfield  {author} {\bibinfo {author} {\bibfnamefont {G.}~\bibnamefont
  {Vaessen}}\ and\ \bibinfo {author} {\bibfnamefont {H.}~\bibnamefont
  {Stein}},\ }\bibfield  {title} {\bibinfo {title} {The applicability of
  catastrophe theory to emulsion phase inversion},\ }\href@noop {} {\bibfield
  {journal} {\bibinfo  {journal} {J. Colloid Interface Sci.}\ }\textbf
  {\bibinfo {volume} {176}},\ \bibinfo {pages} {378} (\bibinfo {year}
  {1995})}\BibitemShut {NoStop}%
\bibitem [{\citenamefont {Dickinson}(1981)}]{Dickinson1981}%
  \BibitemOpen
  \bibfield  {author} {\bibinfo {author} {\bibfnamefont {E.}~\bibnamefont
  {Dickinson}},\ }\bibfield  {title} {\bibinfo {title} {Interpretation of
  emulsion phase inversion as a cusp catastrophe},\ }\href@noop {} {\bibfield
  {journal} {\bibinfo  {journal} {J. Colloid Interface Sci.}\ }\textbf
  {\bibinfo {volume} {84}},\ \bibinfo {pages} {284} (\bibinfo {year}
  {1981})}\BibitemShut {NoStop}%
\bibitem [{\citenamefont {Salager}(1988)}]{Salager1988}%
  \BibitemOpen
  \bibfield  {author} {\bibinfo {author} {\bibfnamefont {J.-L.}\ \bibnamefont
  {Salager}},\ }\bibfield  {title} {\bibinfo {title} {Phase transformation and
  emulsion inversion on the basis of catastrophe theory},\ }\href@noop {}
  {\bibfield  {journal} {\bibinfo  {journal} {P. Becher (Ed.), Encyclopedia of
  Emulsion Science}\ ,\ \bibinfo {pages} {79}} (\bibinfo {year}
  {1988})}\BibitemShut {NoStop}%
\bibitem [{\citenamefont {Zeeman}(1976)}]{Zeeman1976}%
  \BibitemOpen
  \bibfield  {author} {\bibinfo {author} {\bibfnamefont {E.}~\bibnamefont
  {Zeeman}},\ }\bibfield  {title} {\bibinfo {title} {Catastrophe theory},\
  }\href@noop {} {\bibfield  {journal} {\bibinfo  {journal} {Scientific
  American}\ ,\ \bibinfo {pages} {65}} (\bibinfo {year} {1976})}\BibitemShut
  {NoStop}%
\bibitem [{\citenamefont {Vaessen}\ \emph {et~al.}(1996)\citenamefont
  {Vaessen}, \citenamefont {Visschers},\ and\ \citenamefont
  {H.N.}}]{Vaessen1996}%
  \BibitemOpen
  \bibfield  {author} {\bibinfo {author} {\bibfnamefont {G.}~\bibnamefont
  {Vaessen}}, \bibinfo {author} {\bibfnamefont {M.}~\bibnamefont {Visschers}},\
  and\ \bibinfo {author} {\bibfnamefont {S.}~\bibnamefont {H.N.}},\ }\bibfield
  {title} {\bibinfo {title} {Predicting catastrophic phase inversion on the
  basis of droplet coalescence kinetics},\ }\href@noop {} {\bibfield  {journal}
  {\bibinfo  {journal} {Langmuir}\ }\textbf {\bibinfo {volume} {12}},\ \bibinfo
  {pages} {875} (\bibinfo {year} {1996})}\BibitemShut {NoStop}%
\bibitem [{\citenamefont {Bouchama}\ \emph {et~al.}(2003)\citenamefont
  {Bouchama}, \citenamefont {van Aken}, \citenamefont {Autin},\ and\
  \citenamefont {Koper}}]{Bouchama2003}%
  \BibitemOpen
  \bibfield  {author} {\bibinfo {author} {\bibfnamefont {F.}~\bibnamefont
  {Bouchama}}, \bibinfo {author} {\bibfnamefont {G.}~\bibnamefont {van Aken}},
  \bibinfo {author} {\bibfnamefont {A.}~\bibnamefont {Autin}},\ and\ \bibinfo
  {author} {\bibfnamefont {G.}~\bibnamefont {Koper}},\ }\bibfield  {title}
  {\bibinfo {title} {On the mechanism of catastrophic phase inversion in
  emulsions},\ }\href@noop {} {\bibfield  {journal} {\bibinfo  {journal}
  {Colloids Surf. A: Physicochem. Eng. Asp.}\ }\textbf {\bibinfo {volume}
  {231}},\ \bibinfo {pages} {11} (\bibinfo {year} {2003})}\BibitemShut
  {NoStop}%
\bibitem [{\citenamefont {Kolmogorov}(1949)}]{kolmogorov1949droplet}%
  \BibitemOpen
  \bibfield  {author} {\bibinfo {author} {\bibfnamefont {A.}~\bibnamefont
  {Kolmogorov}},\ }\bibfield  {title} {\bibinfo {title} {On droplet breaking-up
  in a turbulent flow},\ }\bibfield  {booktitle} {\emph {\bibinfo {booktitle}
  {Doklady Akad. Nauk. SSSR}},\ }\href@noop {} {\ \textbf {\bibinfo {volume}
  {66}},\ \bibinfo {pages} {825} (\bibinfo {year} {1949})}\BibitemShut
  {NoStop}%
\bibitem [{\citenamefont {Hinze}(1955)}]{hinze1955fundamentals}%
  \BibitemOpen
  \bibfield  {author} {\bibinfo {author} {\bibfnamefont {J.~O.}\ \bibnamefont
  {Hinze}},\ }\bibfield  {title} {\bibinfo {title} {Fundamentals of the
  hydrodynamic mechanism of splitting in dispersion processes},\ }\href@noop {}
  {\bibfield  {journal} {\bibinfo  {journal} {AIChE journal}\ }\textbf
  {\bibinfo {volume} {1}},\ \bibinfo {pages} {289} (\bibinfo {year}
  {1955})}\BibitemShut {NoStop}%
\bibitem [{\citenamefont {Coulaloglou}\ and\ \citenamefont
  {Tavlarides}(1977)}]{Coulaloglou1977}%
  \BibitemOpen
  \bibfield  {author} {\bibinfo {author} {\bibfnamefont {C.}~\bibnamefont
  {Coulaloglou}}\ and\ \bibinfo {author} {\bibfnamefont {L.}~\bibnamefont
  {Tavlarides}},\ }\bibfield  {title} {\bibinfo {title} {Description of
  interaction processes in agitated liquid-liquid dispersions},\ }\href@noop {}
  {\bibfield  {journal} {\bibinfo  {journal} {Chem. Eng. Sci.}\ }\textbf
  {\bibinfo {volume} {32}},\ \bibinfo {pages} {1289} (\bibinfo {year}
  {1977})}\BibitemShut {NoStop}%
\bibitem [{\citenamefont {Taylor}(1932)}]{taylor1932viscosity}%
  \BibitemOpen
  \bibfield  {author} {\bibinfo {author} {\bibfnamefont {G.~I.}\ \bibnamefont
  {Taylor}},\ }\bibfield  {title} {\bibinfo {title} {The viscosity of a fluid
  containing small drops of another fluid},\ }\href@noop {} {\bibfield
  {journal} {\bibinfo  {journal} {Proceedings of the Royal Society of London.
  Series A, Containing papers of a mathematical and physical character}\
  }\textbf {\bibinfo {volume} {138}},\ \bibinfo {pages} {41} (\bibinfo {year}
  {1932})}\BibitemShut {NoStop}%
\bibitem [{\citenamefont {Girotto}\ \emph {et~al.}(2024)\citenamefont
  {Girotto}, \citenamefont {Scagliarini}, \citenamefont {Benzi},\ and\
  \citenamefont {Toschi}}]{girotto2024lagrangian}%
  \BibitemOpen
  \bibfield  {author} {\bibinfo {author} {\bibfnamefont {I.}~\bibnamefont
  {Girotto}}, \bibinfo {author} {\bibfnamefont {A.}~\bibnamefont
  {Scagliarini}}, \bibinfo {author} {\bibfnamefont {R.}~\bibnamefont {Benzi}},\
  and\ \bibinfo {author} {\bibfnamefont {F.}~\bibnamefont {Toschi}},\
  }\bibfield  {title} {\bibinfo {title} {Lagrangian statistics of concentrated
  emulsions},\ }\href@noop {} {\bibfield  {journal} {\bibinfo  {journal} {J.
  Fluid Mech.}\ }\textbf {\bibinfo {volume} {986}},\ \bibinfo {pages} {A33}
  (\bibinfo {year} {2024})}\BibitemShut {NoStop}%
\bibitem [{\citenamefont {Bakhuis}\ \emph {et~al.}(2021)\citenamefont
  {Bakhuis}, \citenamefont {Ezeta}, \citenamefont {Bullee}, \citenamefont
  {Marin}, \citenamefont {Lohse}, \citenamefont {Sun},\ and\ \citenamefont
  {Huisman}}]{bakhuis2021catastrophic}%
  \BibitemOpen
  \bibfield  {author} {\bibinfo {author} {\bibfnamefont {D.}~\bibnamefont
  {Bakhuis}}, \bibinfo {author} {\bibfnamefont {R.}~\bibnamefont {Ezeta}},
  \bibinfo {author} {\bibfnamefont {P.~A.}\ \bibnamefont {Bullee}}, \bibinfo
  {author} {\bibfnamefont {A.}~\bibnamefont {Marin}}, \bibinfo {author}
  {\bibfnamefont {D.}~\bibnamefont {Lohse}}, \bibinfo {author} {\bibfnamefont
  {C.}~\bibnamefont {Sun}},\ and\ \bibinfo {author} {\bibfnamefont {S.~G.}\
  \bibnamefont {Huisman}},\ }\bibfield  {title} {\bibinfo {title} {Catastrophic
  phase inversion in {{high-Reynolds-number}} turbulent {{Taylor-Couette}}
  flow},\ }\href@noop {} {\bibfield  {journal} {\bibinfo  {journal} {Phys. Rev.
  Lett.}\ }\textbf {\bibinfo {volume} {126}},\ \bibinfo {pages} {064501}
  (\bibinfo {year} {2021})}\BibitemShut {NoStop}%
\bibitem [{\citenamefont {Piela}\ \emph {et~al.}(2009)\citenamefont {Piela},
  \citenamefont {Ooms},\ and\ \citenamefont
  {Sengers}}]{piela2009phenomenological}%
  \BibitemOpen
  \bibfield  {author} {\bibinfo {author} {\bibfnamefont {K.}~\bibnamefont
  {Piela}}, \bibinfo {author} {\bibfnamefont {G.}~\bibnamefont {Ooms}},\ and\
  \bibinfo {author} {\bibfnamefont {J.}~\bibnamefont {Sengers}},\ }\bibfield
  {title} {\bibinfo {title} {Phenomenological description of phase inversion},\
  }\href@noop {} {\bibfield  {journal} {\bibinfo  {journal} {Phys. Rev. E}\
  }\textbf {\bibinfo {volume} {79}},\ \bibinfo {pages} {021403} (\bibinfo
  {year} {2009})}\BibitemShut {NoStop}%
\bibitem [{\citenamefont {Binder}(1987)}]{Binder1987}%
  \BibitemOpen
  \bibfield  {author} {\bibinfo {author} {\bibfnamefont {K.}~\bibnamefont
  {Binder}},\ }\bibfield  {title} {\bibinfo {title} {Theory of first-order
  phase transitions},\ }\href@noop {} {\bibfield  {journal} {\bibinfo
  {journal} {Rep. Prog. Phys.}\ }\textbf {\bibinfo {volume} {50}},\ \bibinfo
  {pages} {783} (\bibinfo {year} {1987})}\BibitemShut {NoStop}%
\bibitem [{\citenamefont {Krapivsky}\ \emph {et~al.}(2010)\citenamefont
  {Krapivsky}, \citenamefont {Redner},\ and\ \citenamefont
  {Ben-Naim}}]{Krapivsky2010}%
  \BibitemOpen
  \bibinfo {editor} {\bibfnamefont {P.}~\bibnamefont {Krapivsky}}, \bibinfo
  {editor} {\bibfnamefont {S.}~\bibnamefont {Redner}},\ and\ \bibinfo {editor}
  {\bibfnamefont {E.}~\bibnamefont {Ben-Naim}},\ eds.,\ \href@noop {} {\emph
  {\bibinfo {title} {A kinetic view of Statistical Physics}}}\ (\bibinfo
  {publisher} {Cambridge University Press},\ \bibinfo {year}
  {2010})\BibitemShut {NoStop}%
\bibitem [{\citenamefont {Maa\ss}\ and\ \citenamefont
  {Kraume}(2012)}]{Maass2012}%
  \BibitemOpen
  \bibfield  {author} {\bibinfo {author} {\bibfnamefont {S.}~\bibnamefont
  {Maa\ss}}\ and\ \bibinfo {author} {\bibfnamefont {M.}~\bibnamefont
  {Kraume}},\ }\bibfield  {title} {\bibinfo {title} {Determination of breakage
  rates using single drop experiments},\ }\href@noop {} {\bibfield  {journal}
  {\bibinfo  {journal} {Chem. Eng. Sci.}\ }\textbf {\bibinfo {volume} {70}},\
  \bibinfo {pages} {146} (\bibinfo {year} {2012})}\BibitemShut {NoStop}%
\bibitem [{\citenamefont {Janoschek}\ \emph {et~al.}(2015)\citenamefont
  {Janoschek}, \citenamefont {Toschi},\ and\ \citenamefont
  {Harting}}]{Brilliantov2015}%
  \BibitemOpen
  \bibfield  {author} {\bibinfo {author} {\bibfnamefont {F.}~\bibnamefont
  {Janoschek}}, \bibinfo {author} {\bibfnamefont {F.}~\bibnamefont {Toschi}},\
  and\ \bibinfo {author} {\bibfnamefont {J.}~\bibnamefont {Harting}},\
  }\bibfield  {title} {\bibinfo {title} {Size distribution of particles in
  saturn’s rings from aggregation and fragmentation},\ }\href@noop {}
  {\bibfield  {journal} {\bibinfo  {journal} {Proc. Natl. Acad. Sci. USA}\
  }\textbf {\bibinfo {volume} {112}},\ \bibinfo {pages} {9536} (\bibinfo {year}
  {2015})}\BibitemShut {NoStop}%
\bibitem [{\citenamefont {Fruchart}\ \emph {et~al.}(2021)\citenamefont
  {Fruchart}, \citenamefont {Hanai}, \citenamefont {Littlewood},\ and\
  \citenamefont {Vitelli}}]{Fruchart2021}%
  \BibitemOpen
  \bibfield  {author} {\bibinfo {author} {\bibfnamefont {M.}~\bibnamefont
  {Fruchart}}, \bibinfo {author} {\bibfnamefont {R.}~\bibnamefont {Hanai}},
  \bibinfo {author} {\bibfnamefont {P.}~\bibnamefont {Littlewood}},\ and\
  \bibinfo {author} {\bibfnamefont {V.}~\bibnamefont {Vitelli}},\ }\bibfield
  {title} {\bibinfo {title} {Non-reciprocal phase transitions},\ }\href@noop {}
  {\bibfield  {journal} {\bibinfo  {journal} {Nature}\ }\textbf {\bibinfo
  {volume} {592}},\ \bibinfo {pages} {363} (\bibinfo {year}
  {2021})}\BibitemShut {NoStop}%
\bibitem [{\citenamefont {Grossmann}\ \emph {et~al.}(2016)\citenamefont
  {Grossmann}, \citenamefont {Lohse},\ and\ \citenamefont
  {Sun}}]{Grossmann2016}%
  \BibitemOpen
  \bibfield  {author} {\bibinfo {author} {\bibfnamefont {S.}~\bibnamefont
  {Grossmann}}, \bibinfo {author} {\bibfnamefont {D.}~\bibnamefont {Lohse}},\
  and\ \bibinfo {author} {\bibfnamefont {C.}~\bibnamefont {Sun}},\ }\bibfield
  {title} {\bibinfo {title} {High–reynolds number taylor-couette
  turbulence},\ }\href@noop {} {\bibfield  {journal} {\bibinfo  {journal}
  {Annu. Rev. Fluid Mech.}\ }\textbf {\bibinfo {volume} {48}},\ \bibinfo
  {pages} {53} (\bibinfo {year} {2016})}\BibitemShut {NoStop}%
\bibitem [{\citenamefont {Yi}\ \emph {et~al.}(2021)\citenamefont {Yi},
  \citenamefont {Toschi},\ and\ \citenamefont {Sun}}]{yi2021global}%
  \BibitemOpen
  \bibfield  {author} {\bibinfo {author} {\bibfnamefont {L.}~\bibnamefont
  {Yi}}, \bibinfo {author} {\bibfnamefont {F.}~\bibnamefont {Toschi}},\ and\
  \bibinfo {author} {\bibfnamefont {C.}~\bibnamefont {Sun}},\ }\bibfield
  {title} {\bibinfo {title} {Global and local statistics in turbulent
  emulsions},\ }\href@noop {} {\bibfield  {journal} {\bibinfo  {journal} {J.
  Fluid Mech.}\ }\textbf {\bibinfo {volume} {912}},\ \bibinfo {pages} {A13}
  (\bibinfo {year} {2021})}\BibitemShut {NoStop}%
\bibitem [{\citenamefont {Benzi}\ \emph {et~al.}(1992)\citenamefont {Benzi},
  \citenamefont {Succi},\ and\ \citenamefont {Vergassola}}]{BENZI1992145}%
  \BibitemOpen
  \bibfield  {author} {\bibinfo {author} {\bibfnamefont {R.}~\bibnamefont
  {Benzi}}, \bibinfo {author} {\bibfnamefont {S.}~\bibnamefont {Succi}},\ and\
  \bibinfo {author} {\bibfnamefont {M.}~\bibnamefont {Vergassola}},\ }\bibfield
   {title} {\bibinfo {title} {The lattice boltzmann equation: theory and
  applications},\ }\href@noop {} {\bibfield  {journal} {\bibinfo  {journal}
  {Physics Reports}\ }\textbf {\bibinfo {volume} {222}},\ \bibinfo {pages}
  {145} (\bibinfo {year} {1992})}\BibitemShut {NoStop}%
\bibitem [{\citenamefont {Succi}(2018)}]{succi2018lattice}%
  \BibitemOpen
  \bibfield  {author} {\bibinfo {author} {\bibfnamefont {S.}~\bibnamefont
  {Succi}},\ }\href@noop {} {\emph {\bibinfo {title} {The Lattice Boltzmann
  Equation for complex states of flowing matter}}}\ (\bibinfo  {publisher}
  {Oxford University Press},\ \bibinfo {year} {2018})\BibitemShut {NoStop}%
\bibitem [{\citenamefont {Rowlinson}\ and\ \citenamefont
  {Widom}(2002)}]{rowlinson2002molecular}%
  \BibitemOpen
  \bibfield  {author} {\bibinfo {author} {\bibfnamefont {J.}~\bibnamefont
  {Rowlinson}}\ and\ \bibinfo {author} {\bibfnamefont {B.}~\bibnamefont
  {Widom}},\ }\href {https://books.google.it/books?id=U0WkBfSQudAC} {\emph
  {\bibinfo {title} {Molecular Theory of Capillarity}}},\ Dover books on
  chemistry\ (\bibinfo  {publisher} {Dover Publications},\ \bibinfo {year}
  {2002})\BibitemShut {NoStop}%
\bibitem [{\citenamefont {Bergeron}(1999)}]{bergeron1999forces}%
  \BibitemOpen
  \bibfield  {author} {\bibinfo {author} {\bibfnamefont {V.}~\bibnamefont
  {Bergeron}},\ }\bibfield  {title} {\bibinfo {title} {Forces and structure in
  thin liquid soap films},\ }\href@noop {} {\bibfield  {journal} {\bibinfo
  {journal} {J. Phys.: Condens. Matter}\ }\textbf {\bibinfo {volume} {11}},\
  \bibinfo {pages} {R215} (\bibinfo {year} {1999})}\BibitemShut {NoStop}%
\bibitem [{\citenamefont {Sbragaglia}\ \emph {et~al.}(2012)\citenamefont
  {Sbragaglia}, \citenamefont {Benzi}, \citenamefont {Bernaschi},\ and\
  \citenamefont {Succi}}]{sbragaglia2012supramolecular}%
  \BibitemOpen
  \bibfield  {author} {\bibinfo {author} {\bibfnamefont {M.}~\bibnamefont
  {Sbragaglia}}, \bibinfo {author} {\bibfnamefont {R.}~\bibnamefont {Benzi}},
  \bibinfo {author} {\bibfnamefont {M.}~\bibnamefont {Bernaschi}},\ and\
  \bibinfo {author} {\bibfnamefont {S.}~\bibnamefont {Succi}},\ }\bibfield
  {title} {\bibinfo {title} {The emergence of supramolecular forces from
  lattice kinetic models of non-ideal fluids: applications to the rheology of
  soft glassy materials},\ }\href {https://doi.org/10.1039/C2SM26167G}
  {\bibfield  {journal} {\bibinfo  {journal} {Soft Matter}\ }\textbf {\bibinfo
  {volume} {8}},\ \bibinfo {pages} {10773} (\bibinfo {year}
  {2012})}\BibitemShut {NoStop}%
\bibitem [{\citenamefont {Biferale}\ \emph {et~al.}(2011)\citenamefont
  {Biferale}, \citenamefont {Perlekar}, \citenamefont {Sbragaglia},
  \citenamefont {Srivastava},\ and\ \citenamefont {Toschi}}]{Biferale_2011}%
  \BibitemOpen
  \bibfield  {author} {\bibinfo {author} {\bibfnamefont {L.}~\bibnamefont
  {Biferale}}, \bibinfo {author} {\bibfnamefont {P.}~\bibnamefont {Perlekar}},
  \bibinfo {author} {\bibfnamefont {M.}~\bibnamefont {Sbragaglia}}, \bibinfo
  {author} {\bibfnamefont {S.}~\bibnamefont {Srivastava}},\ and\ \bibinfo
  {author} {\bibfnamefont {F.}~\bibnamefont {Toschi}},\ }\bibfield  {title}
  {\bibinfo {title} {A lattice boltzmann method for turbulent emulsions},\
  }\href {https://doi.org/10.1088/1742-6596/318/5/052017} {\bibfield  {journal}
  {\bibinfo  {journal} {J. Phys. Conf. Ser.}\ }\textbf {\bibinfo {volume}
  {318}},\ \bibinfo {pages} {052017} (\bibinfo {year} {2011})}\BibitemShut
  {NoStop}%
\bibitem [{\citenamefont {Perlekar}\ \emph {et~al.}(2012)\citenamefont
  {Perlekar}, \citenamefont {Biferale}, \citenamefont {Sbragaglia},
  \citenamefont {Srivastava},\ and\ \citenamefont
  {Toschi}}]{perlekar2012droplet}%
  \BibitemOpen
  \bibfield  {author} {\bibinfo {author} {\bibfnamefont {P.}~\bibnamefont
  {Perlekar}}, \bibinfo {author} {\bibfnamefont {L.}~\bibnamefont {Biferale}},
  \bibinfo {author} {\bibfnamefont {M.}~\bibnamefont {Sbragaglia}}, \bibinfo
  {author} {\bibfnamefont {S.}~\bibnamefont {Srivastava}},\ and\ \bibinfo
  {author} {\bibfnamefont {F.}~\bibnamefont {Toschi}},\ }\bibfield  {title}
  {\bibinfo {title} {Droplet size distribution in homogeneous isotropic
  turbulence},\ }\href@noop {} {\bibfield  {journal} {\bibinfo  {journal}
  {Phys. Fluids}\ }\textbf {\bibinfo {volume} {24}},\ \bibinfo {pages} {065101}
  (\bibinfo {year} {2012})}\BibitemShut {NoStop}%
\bibitem [{\citenamefont {Coulaloglou}\ and\ \citenamefont
  {Tavlarides}(1976)}]{Coulaloglou1976}%
  \BibitemOpen
  \bibfield  {author} {\bibinfo {author} {\bibfnamefont {C.}~\bibnamefont
  {Coulaloglou}}\ and\ \bibinfo {author} {\bibfnamefont {L.}~\bibnamefont
  {Tavlarides}},\ }\bibfield  {title} {\bibinfo {title} {Drop size
  distributions and coalescence frequencies of liquid-liquid dispersions in
  flow vessels},\ }\href@noop {} {\bibfield  {journal} {\bibinfo  {journal}
  {AIChE J.}\ }\textbf {\bibinfo {volume} {22}},\ \bibinfo {pages} {289}
  (\bibinfo {year} {1976})}\BibitemShut {NoStop}%
\bibitem [{\citenamefont {Risken}(1996)}]{Risken}%
  \BibitemOpen
  \bibfield  {author} {\bibinfo {author} {\bibfnamefont {H.}~\bibnamefont
  {Risken}},\ }\href@noop {} {\emph {\bibinfo {title} {The Fokker-Planck
  Equation: methods of solution and applications}}}\ (\bibinfo  {publisher}
  {Springer International Publishing},\ \bibinfo {year} {1996})\BibitemShut
  {NoStop}%
\bibitem [{\citenamefont {Yi}\ \emph {et~al.}(2024)\citenamefont {Yi},
  \citenamefont {Girotto}, \citenamefont {Toschi},\ and\ \citenamefont
  {Sun}}]{yi2024divergence}%
  \BibitemOpen
  \bibfield  {author} {\bibinfo {author} {\bibfnamefont {L.}~\bibnamefont
  {Yi}}, \bibinfo {author} {\bibfnamefont {I.}~\bibnamefont {Girotto}},
  \bibinfo {author} {\bibfnamefont {F.}~\bibnamefont {Toschi}},\ and\ \bibinfo
  {author} {\bibfnamefont {C.}~\bibnamefont {Sun}},\ }\bibfield  {title}
  {\bibinfo {title} {Divergence of critical fluctuations on approaching
  catastrophic phase inversion in turbulent emulsions},\ }\href@noop {}
  {\bibfield  {journal} {\bibinfo  {journal} {Phys. Rev. Lett.}\ }\textbf
  {\bibinfo {volume} {133}},\ \bibinfo {pages} {134001} (\bibinfo {year}
  {2024})}\BibitemShut {NoStop}%
\end{thebibliography}%
\end{document}